\documentclass[a4paper,fleqn,usenatbib]{mnras}  

\usepackage{newtxtext,newtxmath}
\usepackage{hyperref}
\usepackage[T1]{fontenc}
\usepackage{ae,aecompl}
\usepackage{amsmath}
\usepackage[export]{adjustbox}
\newcommand*\diff{\mathop{}\!\mathrm{d}}

\usepackage{pifont}
\usepackage{graphicx}	
\usepackage{amsmath}	
\usepackage{amssymb}	
\usepackage[font=normalsize]{caption}
\usepackage{float}
\usepackage[caption=false]{subfig} 
 \usepackage[flushleft]{threeparttable}
\usepackage{verbatim}
\usepackage{stackengine}

\title[Binary Mass Transfer using Bipolytropes]{Numerical Simulations of Mass Transfer in Binaries with Bipolytropic Components}

\author[ Kadam et al.]{
Kundan Kadam$^{1,2}$\thanks{E-mail: {kundan.kadam@csfk.mta.hu}},
Patrick M. Motl$^{3}$,
Dominic C. Marcello$^{1}$,
Juhan Frank$^{1}$, and
\newauthor
Geoffrey C. Clayton$^{1}$\\
$^{1}$Department\ of Physics \& Astronomy, Louisiana State University, Baton Rouge, LA 70803, USA\\
$^{2}$Konkoly Observatory, Research Centre for Astronomy and Earth Sciences, Hungarian Academy of Sciences, Konkoly-Thege Mikl\'{o}s \\
\'{u}t 15-17, 1121 Budapest, Hungary\\
$^{3}$The School of Sciences, Indiana University Kokomo, Kokomo, Indiana 46904, USA}
\date{Accepted 2018 September 10; Received 2018 September 10; in original form 2017 August 05}

\pubyear{2017}

\begin{document}
\label{firstpage}
\pagerange{\pageref{firstpage}--\pageref{lastpage}}
\maketitle

\begin{abstract}

We present the first self-consistent, three dimensional study of hydrodynamic simulations of mass transfer in binary systems with bipolytropic (composite polytropic) components. 
In certain systems, such as contact binaries or during the common envelope phase, the core-envelope structure of the stars plays an important role in binary interactions.
In this paper, we compare mass transfer simulations of bipolytropic binary systems in order to test the suitability of our numerical tools for investigating the dynamical behaviour of such systems. 
The initial, equilibrium binary models possess a core-envelope structure and are obtained using the bipolytropic self-consistent field technique.
We conduct mass transfer simulations using two independent, fully three-dimensional, 
Eulerian codes - Flow-ER and Octo-tiger.
These hydrodynamic codes are compared across binary systems undergoing unstable as well as stable mass transfer, and the former at two resolutions. 
The initial conditions for each simulation and for each code are chosen to match closely so that the simulations can be used as benchmarks. 
Although there are some key differences, the detailed comparison of the simulations suggests that there is remarkable agreement between the results obtained using the two codes.
This study puts our numerical tools on a secure footing, and enables us to reliably simulate specific mass transfer scenarios of binary systems involving components with a core-envelope structure.

\end{abstract}

\begin{keywords}
binaries: close -- hydrodynamics -- methods: numerical  
\end{keywords}



\section{Introduction}

About half of the stars in the sky are in reality binaries or multiple star systems \citep{Trimble1983}, and
with an expected surge in detection of transient events due to large-scale surveys, 
it becomes more relevant to understand the nature of transient phenomena involving binary interactions. 
In recent years there has been considerable interest in modelling dynamical interactions of binary systems involving main sequence as well as post main sequence stars.
For example, \cite{Nandez2014} conducted a smooth particle hydrodynamics (SPH) study to simulate the evolution of the progenitor binary of V1309 Scorpii, which was a contact binary merger event, through the common envelope phase with a red giant primary and a degenerate as well as a main sequence secondary. 
This project is motivated by our interest in the understanding of the mass transfer events in binary systems, and in particular, 
the evolutionary status, internal structure as well as stability criteria of contact binary systems. 
The contact binaries in equilibrium exhibit many apparently contradictory properties such as Kuiper's paradox (conformity to Roche geometry, \cite{Kuiper1941}), Lucy's paradox (equal surface temperature of unequal mass components, \cite{Lucy1968}) and secular stability over billions of years \citep{LiZhang2007}. 
\citet{Stepien2009} postulates a model for energy transported from the primary to the lower mass component by means of a steadily circulating stream in the equatorial region, in order to resolve Lucy's paradox, while also maintaining thermal stability of the system. 
The dynamic stability and exact nature of the large-scale circulations has not yet been verified from first principles with models which are fully self-consistent. Modelling the components with a core-envelope structure may help elucidate the dynamical stability of these flows in contact binary systems.

With the above issues in the theoretical understanding of contact binary systems in mind, in this paper we describe a set of grid-based numerical tools suitable for studying the hydrodynamic evolution of mass transferring binary systems with each component having a core and an envelope. 
One way to model the internal structure of such a star is using bipolytropes (sometimes also referred to as composite polytropes), where the core and the envelope of the star are allowed to have different polytropic indices \citep{Henrich1941}.
We are investigating the dynamical evolution of binaries involving stars having a core-envelope structure, and, in particular, contact binaries with hydrodynamic simulations in a series of three papers.
In \cite{Kadam2016} (henceforth paper I) we established the viability of the self-consistent field method to produce single rotating bipolytropic structures in equilibrium. In the current paper (paper II) we present a set of Eulerian numerical tools that we have developed for investigating the dynamical phases of mass transfer in bipolytropic binary systems.
 A suite of hydrodynamic simulations of bipolytropic binary systems is conducted in order to test the suitability of our numerical methods for investigating such scenarios.
We plan to focus on contact binary systems, ${\rm e.g.}$ the evolution of the progenitor of V1309 Scorpii, in paper III. 

There are two key steps in order to conduct fully three dimensional hydrodynamic simulations of binary systems. In the first step, we build the initial models using a modified version of Hachisu's self-consistent field technique \citep{Hachisu1986b}, which we call the bipolytropic self-consistent field (BSCF) technique. 
A bipolytrope is a structure with two polytropes stacked on top of each other, so that the core and the envelope have two different equations of state. The composition difference between the core and the envelope can be mimicked by specifying a jump in the molecular weight in the BSCF method.
In the next step, we use the bipolytropic binary systems as quiet initial conditions and study their dynamic evolution using suitable hydrodynamic codes. 
We used two independent, explicit, Eulerian codes -- Flow-ER \citep{Motl2002, DSouza2006} with a fixed cylindrical grid and Octo-tiger \citep{MarcelloOct2017} with a Cartesian AMR (adaptive mesh refinement) grid.
The binary systems were chosen to match closely across the two codes, so that the results could be compared directly and could be used as a benchmark for each code. 
In the first set of simulations the initial mass ratio of the binary systems was set to 0.7, so that with a convective envelope of the donor, the mass transfer would be unstable. These simulations were conducted until merger at two grid resolutions with both the codes. 
In order to compare behaviour of the codes for binaries undergoing stable mass transfer, we also conducted simulations with an initial mass ratio of 0.6 for a reasonably large number of orbits.
We examined the time evolution of key parameters, including separation, mass transfer rate, mass ratio, orbital and spin angular momenta, as well as density cross sections.
The dynamical behaviour of the systems agreed remarkably well and the differences can be accounted for by considering the differences between the two codes, especially the inclusion of shock heating in Octo-tiger and the limited resolution of the stellar cores in Flow-ER.
This study proved the reliability of our numerical techniques and highlighted the suitability of Octo-tiger for investigating dynamical binary interactions with bipolytropic components.

The structure of this paper is as follows. Section \ref{bibi} describes the BSCF method that is used for generating the initial models. In section \ref{numal} we describe the two hydrodynamics codes that are used for the dynamical evolution of the binary systems. Section \ref{tim} explains the setup and properties of the suite of binary systems used as initial conditions for the simulations. 
In section \ref{mts} we describe how the binary parameters are calculated, and then compare and analyse the simulation results for both unstable as well as stable mass transfer scenarios.
In section \ref{summary} we summarize our findings and discuss the applicability of our numerical tools for future investigations.

\section{Bipolytropic Binaries}
\label{bibi}

First introduced by \cite{OstrikerMark1968}, the self-consistent field (SCF) technique is an iterative method of finding the equilibrium solutions of a rotating, gravitationally bound fluid.
This method was further improved in \cite{Hachisu1986a, Hachisu1986b} to produce self-gravitating, rapidly rotating stars as well as systems without axial symmetry, such as binary or multiple star systems. 
This improved method of calculating equilibrium structures is called Hachisu's self-consistent field (HSCF) technique.
In the past, the HSCF technique has been used to construct the initial models of detached, close binary systems with a wide range of mass ratios and Roche lobe filling factors (${\rm e.g.}$ \cite{Motl2002, DSouza2006,Even2009}). The binary systems that were generated had either a single polytropic or a zero-temperature white dwarf equation of state for both the stars, and they were subsequently used as initial conditions for the study of their dynamical behaviour.
In this section we describe a computational method to obtain the equilibrium configuration of close binary systems having a bipolytropic structure for each component.
This is achieved by using the bipolytropic self-consistent field (BSCF) method.
This technique is similar to the method described in paper I, which is used for constructing rapidly rotating single bipolytropic spheroidal as well as toroidal structures in equilibrium.

\subsection{Implementation}
\label{Implementation}
We make the following assumptions while using the BSCF technique for a binary system. 
The stars are assumed to be rotating synchronously in a circular orbit with a uniform angular velocity ($\Omega$) such that the system is in hydrostatic equilibrium in the frame of reference corotating with the system. The basic equations describing the internal structure of each bipolytropic star are similar to those described in paper I. 

The polytropic equation of state assumes a power law relationship between the pressure and the density of the gas-
\begin{equation}
 P=\kappa\rho^{1+\frac{1}{n}}= {\rm \kappa} \rho^{\gamma},
\end{equation}
where $n$ is the polytropic index, ${\rm \kappa}$ is the polytropic constant and $\gamma$ is the polytropic exponent.
A bipolytropic star has one such polytropic index for the core ($n_{\rm c}$) and one for the envelope ($n_{\rm e}$). 
Since the pressure and temperature are continuous functions across the core-envelope interface, this implies-
\begin{equation}
\frac{\rho_{\rm ci}}{\rho_{\rm ei}}=\frac{\mu_{\rm c}}{\mu_{\rm e}}=\alpha,
\end{equation}
where $\rho_{\rm ci}$, $\rho_{\rm ei}$ are the densities at the interface and $\alpha$ is the ratio of average molecular weights between the core and the envelope - $\mu_{\rm c}$ and $\mu_{\rm e}$ respectively. 
In the corotating frame of reference, the steady-state equation of motion for each stellar component, $i$, in a binary can be written as-
\begin{equation}
H + \Phi + \Omega^2 \Psi = {\rm C}_i,
\end{equation}
where $H$ is the specific enthalpy of the gas, $\Phi$ is the gravitational potential and $\Psi$ is the coordinate-dependent part of the centrifugal potential. 
Here ${\rm C_i}$ are the integration constants which are analogous to the Bernoulli constant in classical fluid flows. These constants are different for each polytrope, in general, hence in a bipolytropic binary configuration there are four such constants.
The density and hence the enthalpy is zero at the surface of each star.

\begin{figure} 
\centering
\includegraphics[width=3.3in]{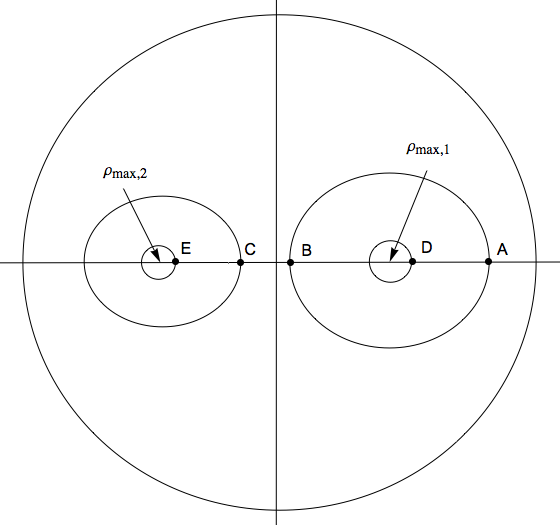}
\caption{A schematic showing the boundary points in the equatorial plane that determine the structure of a bipolytropic binary system. In addition to these boundary conditions, $n_{\rm c}$, $n_{\rm e}$ and $\mu_{\rm c}/ \mu_{\rm e}$ need to be specified for both stars in order to obtain the initial model. }
\label{fig:bibi_schematic}
\end{figure}

The BSCF method for obtaining a binary system requires seven boundary conditions in order to solve the hydrostatic equations for the two stars along with Poisson's equation. These are depicted in Figure \ref{fig:bibi_schematic}.
The boundary conditions are chosen to be the maximum densities of the two stars $\rho_{\rm max, 1}$,  $\rho_{\rm max, 2}$, the location of the outer boundary of star 1 (A), the location of the inner boundary of the same star (B) and the location of the inner boundary of star 2 (C). Here subscript $1$ denotes the star on the right in Figure \ref{fig:bibi_schematic}, so that two of its boundaries are specified. The three points are collinear and lie along the line joining the centre of mass of the two stars. In order to obtain a bipolytropic binary, the locations of the boundary points of the core for both the stars (points D and E) need to be specified. In addition to the above, the polytropic indices of the core and the envelope, as well as the ratio of molecular weights between them for both the stars need to be specified, the choices of which depend on the type of stars one wants to represent.  

An iterative scheme to solve Equation 3, along with Poisson's equation in the BSCF method works as follows.
Note that the integration constants ${\rm C_i}$ can be different for the core and the envelope of each star. 
An initial density distribution satisfying all the outer boundary conditions must be guessed for both the stars. The gravitational potential can be calculated by solving Poisson's equation. The numerical technique to solve this equation is identical to that used during the hydrodynamical evolution, and is described in section \ref{numal} for each code.
Since the density, and hence the enthalpy, at the surface is forced to vanish, the surface boundary conditions at point A, B and C are
\begin{equation}
H(A)= H(B) = H(C)=0.
\end{equation}
The envelope integration constant ${\rm C_{e1}}$ is the same at points A and B, which allows us to calculate the angular velocity of the binary,
\begin{equation}
\Omega^2= -\frac{\Phi_{\rm A}-\Phi_{\rm B}}{\Psi_{\rm A}-\Psi_{\rm B}}.
\end{equation}
When we put this value of angular velocity back in Equation 3 for points A and C, the envelope integration constants for both the stars, ${\rm C_{e1}}$ and ${\rm C_{e2}}$, can be found.

Now we are at the same point as Equation 18 in paper I, where the rotational velocity of the system and the envelope integration constant for each star are known. The new density distribution for each star can be found by following similar steps. 
In the next iteration cycle, the axis of rotation is allowed to move to the location of the centre of mass of the system along the line joining the centre of mass of each star.
The centrifugal potential is calculated with respect to this new axis of rotation.
The computation is performed for two non-overlapping regions for star 1 and 2, such that each star lies completely inside only one of the regions. 
The iteration procedure is repeated until the relative changes in ${\rm C_{e1}}$, ${\rm C_{e2}}$, ${\rm H_{\rm max, 1}}$, ${\rm H_{\rm max, 1}}$ and $\Omega$ are smaller than a prescribed convergence criterion, $\delta$.

The quality of a converged solution is quantified using the virial error 
\begin{equation}
VE=\frac{ \mid 2T+W+3\Pi \mid} { \mid W \mid},
\end{equation}
where T is the total kinetic energy, W is the total gravitational energy and $\Pi$ is the volume integral of the pressure,
\begin{equation}
\Pi = \int_V P \diff{V}.
\end{equation}
VE is a global measure of the accuracy of the equilibrium of a non-linear dynamical system and should ideally be zero. 
The calculations are performed in dimensionless form of the variables, or ``code units", where the maximum density on the grid (${\rm \rho_{0}}$), the gravitational constant (${\rm G}$) and a suitable length, typically the size of the computational grid, are set to unity.

Similar to the HSCF method, the BSCF technique suffers from one major drawback. Although close binaries with a wide range of mass ratios and core mass fractions can be generated using the BSCF technique, physical quantities of interest such as the mass ratio, Roche lobe filling factors and the core mass fractions cannot be specified a priori or controlled precisely. The control parameters $\rho_{\rm max, 1}$,  $\rho_{\rm max, 2}$ and locations of points A through E need to be adjusted through trial and error, to produce systems which have desired properties. 
Note that there are no analytical solutions for binary bipolytropes (or polytropes) to compare the converged solutions with. However, one way to test the viability of the solutions obtained is to evolve these models with a hydrodynamic code and compare the results with our expectations from well known analytical approximations.

\section{Numerical Algorithms}
\label{numal}
Once an equilibrium binary system is generated using the BSCF method, it can be used as a quiet initial condition and can be evolved numerically using a suitable hydrodynamic code. 
We conduct the binary evolution using two explicit, grid-based (Eulerian), fully three-dimensional, hydrodynamic codes -- Flow-ER and Octo-tiger. This section briefly describes these two computational fluid dynamics algorithms.

\subsection{The Flow-ER Code}
\label{sec:flower}

Flow-ER is an explicit, conservative, finite-volume, three-dimensional Eulerian code, which is second order accurate in both time and space.  The code is similar to the ZEUS code \citep{StoneNorman1992} and is implemented on a uniform cylindrical grid. 
The code tracks the following fluid quantities: mass density ($\rho$), angular momentum density (${\rm A_m}$), vertical momentum density (${\rm T_m}$) and radial momentum density (${\rm S_m}$). 
In addition, the entropy tracer- 
\begin{equation}
 \tau = (\epsilon \rho)^{1/\gamma},
\end{equation} 
where $\epsilon$ is the internal energy per unit mass and $\gamma$ is the ratio of specific heats, is also tracked with an assumption of adiabatic evolution.
The set of differential equations solved correspond to the conservation laws of above four quantities, in addition to the internal energy equation for $\tau$
in a reference frame rotating with a constant angular velocity.

The gravitational potential at each timestep is obtained by numerically solving Poisson's equation with the same technique as described in paper I.
An ideal gas equation of state is used to close the system of equations
 \begin{equation}
P=(\gamma_{\rm evo}-1) \tau^{\gamma_{\rm evo}} = (\gamma_{\rm evo}-1)\rho \epsilon.
\end{equation}
 Here, $\gamma_{\rm evo}$ is the polytropic exponent used while evolving the system in time in the hydrodynamic code. All simulations in this paper are performed using a constant value of $\gamma_{\rm evo} = 1+1/1.5 = 5/3$, so that the fluid is convectively stable.
The outer boundary conditions are outflow and are implemented in such a way that the material is free to move out of the grid, but it is not allowed to move back in from the outermost cells. At the axis of the cylindrical grid, the innermost annulus of cells is averaged in the azimuthal direction at a given layer in the vertical direction.  
For a detailed description of the treatment of the advection terms, source terms, time centreing and numerical tests, see \citet{Motl2002} and \citet{DSouza2006}.

With the BSCF method, the centre of mass is allowed to drift along the line joining the two stars. 
We transport the density field from the BSCF grid to the Flow-ER grid such that the centre of mass coincides with the coordinate axis of the cylindrical grid as closely as possible. The translation is implemented on the pressure field and the density is recalculated at the end, in order to preserve the discontinuity at the interface of the core and the envelope. 
This translation is performed using a bicubic spline interpolation scheme described in \cite{NumFort}. 

Consider a star placed in a cylindrical computational grid.
If the star is placed such that its centre lies at the axis of the grid, the resolution in the azimuthal direction equals that of the grid.
However, if the star is placed away from the axis, as a component of a binary system, its azimuthal resolution decreases sharply and keeps getting worse as it moves farther from the axis. 
The core in a bipolytropic star is especially affected by this degradation of the resolution because of its relatively small size.
Unlike a single bipolytrope placed at the centre of the cylindrical grid, or a polytropic binary, we noticed that hydrodynamic evolution of bipolytropic binary systems in Flow-ER showed a numerical artefact-- the central density (defined as the maximum density of a star) of each bipolytropic component showed a fall of the order of a few per cent on a dynamical timescale.
To investigate this issue further, we conducted a set of simulations with non-interacting, bipolytropic binaries having a mass ratio of unity. 
In conclusion, we found that after a threshold, the fall in the central density was roughly proportional to the absolute slope of the density near its maximum. This implied that if the cores of bipolytropic binaries were not sufficiently resolved, the finite volume method as implemented in Flow-ER with a cylindrical geometry could not preserve steep density gradients near the centre of a bipolytropic star.   
This resulted in numerical diffusion of bipolytropic cores, and was reflected by the fall of the central densities of both the binary components. 
The numerical diffusion can rearrange gas inside a star and affect the dynamical behaviour of the binary.
Increasing the resolution of the Flow-ER simulations, however, is computationally very expensive.
Although this resolution-dependent numerical effect appears to be a major drawback, the study of mass transferring bipolytropic binaries, as discussed in sections \ref{sec:hydro} and \ref{sec:hydro1}, shows that the overall behaviour of a binary system remains unaffected.

\subsection{The Octo-tiger Code}
\label{sec:octotiger}

Octo-tiger is a state of the art, explicit, conservative, hydrodynamic code with a Cartesian AMR grid, developed for modelling mass transfer in binary systems \citep{MarcelloOct2017}.
The AMR uses an octree data structure. Refined nodes have eight regularly spaced children, providing twice the resolution in their respective domains.
The self-similarity in octree structures provides a relatively simple to implement AMR framework, as compared to other alternatives such as patch-based AMR.  
The equations that are numerically solved in Octo-tiger, conservation of mass and momentum as well as the internal energy equation (evolution of $\tau$) are similar to those in Flow-ER. The code also implements conservation of total energy explicitly, in a manner similar to \cite{Marcello2012}.
The ideal gas equation of state used in Octo-tiger is identical to that implemented in the Flow-ER code. However, the energy expression used for the treatment of shocks at high velocities, where the kinetic energy dominates the internal energy of the fluid, is different. In general, the specific internal energy is calculated as $\epsilon = E - v^2/2$, where $E $ is the specific total energy inside a cell and $v$ is the velocity. 
In shocks, the very high velocities involved could make calculations of $\epsilon$ inaccurate, as $E >> \epsilon$. 
Accurate computation of the temperature is impossible in this case, since the pressure is proportional to the internal energy. Octo-tiger uses the dual energy formalism of \cite{Bryan1995} to address this issue. 
The treatment of the boundary in Octo-tiger similar to that in Flow-ER and the material is allowed to flow out but not back into the grid from the outermost boundary points.  
Thus Flow-ER evolves fluids adiabatically with an ideal gas equation of state, while Octo-tiger takes shock heating into account and uses the dual energy formalism with an ideal gas equation of state.

Grid-based hydrodynamic codes using finite volume methods typically conserve linear momentum or angular momentum, but not both.
Octo-tiger solves Poisson's equation using a modified Fast Multipole Method (FMM)  with complexity $\mathcal{O}(N)$ where the potential is computed as a multipole expansion to third order. Octo-tiger implements a modified version of the FMM of \cite{Dehnen2000} which conserves both linear and angular momenta to machine precision \citep{Marcello2017}. The hydrodynamic module also conserves linear and angular momenta to machine precision through the novel technique of \cite{DL2014}.
The conservation of mass and total energy as well as both linear and angular momenta to machine precision is important for investigating long term, dynamical stability of an evolving binary system.

Octo-tiger is implemented in ${\rm C \texttt{++}}$ and parallelized using the High Performance ParalleX (HPX) runtime system developed in collaboration with the Ste$||$ar\footnote{\href{http://stellar.cct.lsu.edu/about/}{http://stellar.cct.lsu.edu/about/}} (Systems Technology, Emergent Parallelism, and Algorithm Research) group at the Center for Computation \& Technology (CCT) at the Louisiana State University \citep{HPX}. 
HPX allows efficient use of computational resources through better scaling and performance as compared to more conventional programming models such as MPI (Message Passing Interface).

\section{The Initial Models}
\label{tim}

In this section we describe the initial models for the suite of bipolytropic binary simulations conducted using the two codes, Flow-ER and Octo-tiger. 
We conducted a total of six benchmark simulations (see Table \ref{table:simpar}). A set of four simulations started with an initial mass ratio $q_0 \approx 0.7$, are called Q0.7; while two additional simulations started with an initial mass ratio $q_0 \approx 0.6$ are called Q0.6.
The properties of the initial models were matched closely for a given mass ratio.

The choice of the internal structure of the initial models was affected by the limitations of the code Flow-ER, as well as our intention of testing the numerical tools with bipolytropic binaries, as opposed to simulating a real evolutionary scenario. 
Both Q0.7 and Q0.6 models were constructed such that the lower mass, secondary component will overflow its Roche lobe.
As explained in section \ref{sec:flower}, in the case of Flow-ER simulations the steep density gradients near a star's centre result in a proportional numerical diffusion of the core over a dynamical time.
Thus for the purpose of benchmarking, we chose moderate values for the core polytropic indices of the bipolytropic components. 
The core and envelope polytropic indices for all stars in the simulations were $(n_c, n_e)=(3,3/2)$. These values are suitable for a radiative core and a convective envelope.
Since the gradient of the central density also increases with the increasing $\alpha$, we selected a value of $\alpha=2$ for all stars.
This value is close to the ratio of molecular weights between fully ionized helium core and a solar composition envelope ($\alpha = 2.22$).
 We would like to emphasize that this internal structure of the bipolytropes is not fully consistent with what is expected from a real evolutionary scenario.
However, the detailed study of simulations with bipolytropic binary components having deep convective envelopes is useful for code testing and comparison.
Such benchmarking is essential so that the novel behaviour of simulations, with stars having realistic internal structure, can be trusted in the future.

Table \ref{table:simpar} lists the resolutions used to carry out the benchmark simulations. 
The prefix ``cyl" or ``oct" denotes the code used (Flow-ER or Octo-tiger respectively) and postfix ``LR" or ``HR" denotes the use of low or high resolution respectively.
The cylLR model had a grid resolution of ${\rm NUMR \times NUMZ \times NUMPHI = 258 \times 130 \times 256}$. In the cylHR model, the resolution of each star was doubled in the ${\rm R}$-direction, and approximately doubled in the other two directions. 
The total resolution of cylHR model was ${\rm NUMR \times NUMZ \times NUMPHI = 386 \times 258 \times 512}$.
In the Flow-ER grid, the mesh interval in the ${\rm R}$-direction was set to that in the ${\rm z}$-direction, ${\rm i.e.}$ ${\rm d R}={\rm d z}$. 
The domain of cylHR simulation was smaller in the radial direction as compared to that of cylLR, and contained only about 58\% of the volume. 
The AMR grid in Octo-tiger changes dynamically according to the matter distribution, hence we specify only the initial effective resolution of each simulation. 
The resolution criteria in Octo-tiger are based on the total number of refinement levels specified as well as the density distribution. 
The core of both the stars were resolved to the finest refinement level, and both the envelopes are resolved to the next level. 
The rest of the grid was refined such that each cell has approximately an equal amount of mass.
The initial effective resolution for an Octo-tiger simulation was obtained by multiplying the total number of nodes by the number of subgrids ($8^3$), which equals the total number of cells on the grid.
The model in octHR had two more levels of refinement as compared to octLR and its computational domain was also twice as large (eight times more volume), which effectively doubled the resolution of the cores of both the donor and the accretor.
For comparison, the core resolutions are listed in Table \ref{table:simpar}. In the case of the Flow-ER code, the core resolution is specified as the number of cells in the radial and the azimuthal direction, and in the case of Octo-tiger, the core resolution is specified as the number of cells in the X and Y direction. 
Distances in all four simulations were normalized with respect to the radius of the BSCF grid in the cylLR model. Note that the size of the computational grid in a simulation can be different from that of the corresponding BSCF model. 
The sizes of the computational domains for the Q0.7 simulations are also listed in Table \ref{table:simpar} in code units.

\begin{table*}
	\begin{center}
	\caption{Simulation Resolutions }
		\label{table:simpar}
		\vspace{0.2cm}
	\begin{tabular}{lllllllll} 
		\hline
		Benchmark & Model ID &  { Resolution \ding{192} }&   { Domain Size  \ding{193} } & { Core Resolution  \ding{194} } &  Notes     \\
		 Set &    &       &    & Donor   \hspace{0.5cm}     Accretor  &      \\
		\hline
		 Q0.7 & cylLR & $258 \times 130 \times 256 \approx 8.6$ M &  $ {1.9}\times0.96$ & $15\times11 \hspace{1cm}  17\times {9}$      \\
		& cylHR & $386 \times 258 \times 512 \approx 51$ M & $ {1.4}\times0.96$ & $30\times14 \hspace{1cm} 34\times22 $ \\
		& octLR & $7721 \times 8^3 \approx 3.9 $ M &  $8.0\times8.0\times8.0$  &$32\times32  \hspace{1cm}    34\times35$  &  8 levels of refinement    \\
		& octHR & $ 32841 \times 8^3 \approx 17$ M & $16\times16\times16$  &$ 64\times64 \hspace{1cm}  68\times68 $ & 10 levels of refinement      \\
		 \\
 		{ Q0.6 }& { cylEE }& $258 \times 130 \times 256 \approx 8.6$ M &  $ {1.9}\times0.96$ & $15\times9 \hspace{1.1cm}  17\times 9$      \\
		& { octEE }& $6300 \times 8^3 \approx 3.2 $ M &  $8.0\times8.0\times8.0$  &$32\times32  \hspace{1cm}    36\times36$  &  8 levels of refinement    \\	 
                 \hline
	\end{tabular}
	\end{center}
	\begin{tablenotes}
               \item  {\ding{192}} The resolution in the cylindrical grid is specified as ${\rm NUMR \times NUMZ \times NUMPHI}$. The initial resolution in the AMR grid is specified as total number of grid cells obtained by multiplying the total number of nodes by $8^3$ subgrids. \\
		 {\ding{193}} The sizes of the computational domains are specified in code units. For the Flow-ER simulations the sizes are in $R$ and $z$ directions and for the Octo-tiger the sizes are in $X$, $Y$ and $Z$ direction.\\
                {\ding{194}} The core resolution for cylindrical grid is in terms of the number of cells in $R$ and $\phi$ direction, and that for AMR grid is the number of cells in $X$ and $Y$ direction.
         \end{tablenotes}	
\end{table*}

{  
The setup for the Q0.6 simulations was similar to the Q0.7 simulations at the low resolution, ${\rm i.e.}$ the ``LR" models. 
The postfix ``EE" in Table \ref{table:simpar} denotes our intention of an extended evolution, in order to demonstrate that stable mass transfer can be handled adequately by our codes. 
Thus cylEE model had a grid resolution of ${\rm NUMR \times NUMZ \times NUMPHI = 258 \times 130 \times 256}$ in Flow-ER, while the corresponding model octEE had 8 levels of refinement.
The internal structure of both the components was also similar to the Q0.7 models, ${\rm i.e.}$ $(n_c, n_e)=(3,3/2)$ and $\alpha=2$. 
}

 The initial values of the binary system parameters {for the benchmark simulations} are listed in Table \ref{table:initbibicode} in code units. Here a subscript $\rm D$ denotes the donor, $\rm A$ denotes the accretor, $\rm c$ denotes the core and $\rm e$ denotes the envelope. The quantities listed in this table are-- mass ($M$), effective radius ($R$), maximum density ($\rho_{\rm max}$), core mass fraction $(M_{\rm c}/M)$, {rotational (spin) angular momentum $J$}, polytropic constant of the core ($K_{\rm c}$) and the envelope ($K_{\rm e}$) for each of the components, along with the mass ratio ($q$), the Roche lobe filling factor by volume (Rlff), separation ($a$), period ($P$), {orbital angular momentum $J_{\rm orb}$}, virial error (VE, see Equation 6) and the minimum density on the grid at the beginning of simulation ($\rho_{\rm min }$). The Rlff is defined in terms of the ratio of the volume of the component above a density threshold of $10^{-5}$ to the volume of its Roche lobe.
{
Although for a given mass ratio the constructed initial models closely match each other, one may notice that they are not identical. } 
The accuracy of the converged solution depends on the resolution and the departure from equilibrium should tend to converge to zero with the increase in resolution. Thus even if the input parameters are precisely specified via the locations of the five boundary points specified in the BSCF method, obtaining identical models for this study is not possible. 
Based on our previous experience with mass transferring binary systems with polytropic components \citep{Motl2017}, this slight difference in the initial models should not affect the results.
{ Assuming a primary with solar mass and radius, the properties of the initial binary can be calculated in solar units, which are listed in Table \ref{table:initbibi}. 
Such binary systems involving low-mass stars ({approximately} $0.5-1.3 M_{\odot} $) are thought to remove angular momentum through magnetic braking in order to form a semi-detached or contact configuration \citep{Kawaler1988}.}

\begin{table*}
	\begin{center}
	\caption{Initial bipolytropic binary system parameters in code units }
		\label{table:initbibicode}
		\vspace{0.2cm}
	\begin{tabular}{lllllll} 
		\hline
		 {Parameter \ding{192}} / Model ID  &  cylLR  &  cylHR &   octLR &  octHR  &  {cylEE }&  {octEE }\\
		\hline
		\bf $q$ & $0.7022$ & $0.6971$ & $0.6765$ &   $0.6936$ &   $0.5979$ &   $0.5905$ \\
		\bf $M_{\rm D}$ &  $5.702 \times 10^{-3}$ & $5.485\times 10^{-3}$ & $5.468 \times 10^{-3}$ & $  5.456\times 10^{-3}$ & $  4.047\times 10^{-3}$ & $  4.008\times 10^{-3}$  \\
		\bf $R_{\rm D}$ & $0.3493$ &  $0.3502$  & $0.3496$ &$ 0.3492 $ &$ 0.3383 $ &$ 0.3359 $  \\
		\bf $\rho_{\rm max,D}$ & $0.8140$ & $0.8360$ & $0.8140$ & $0.8140$ & $1.000$ & $1.000$   \\
		\bf $(M_{\rm c}/M)_{\rm D}$ & $8.263 \times 10^{-2} $ & $9.784 \times 10^{-2} $ & $9.920 \times 10^{-2}  $ &  $9.857 \times 10^{-2} $  &  $0.1787  $  &  $0.1808 $  \\
		\bf ${\rm J_D}$ & $1.317\times 10^{-5} $& $1.217 \times 10^{-5} $ & $1.226\times 10^{-5}$ &$ 1.282 \times 10^{-5}$  &$ 6.858 \times 10^{-6}$  &$ 6.747 \times 10^{-6}$ \\
		\bf $K_{\rm c,D}$ & $6.595 \times 10^{-3}$& $6.519\times 10^{-3}$  &  $6.476\times 10^{-3}$ & $ 6.621\times 10^{-3}$ & $ 5.761\times 10^{-3}$ & $ 5.712\times 10^{-3}$  \\
		\bf $K_{\rm e,D}$ & $2.807\times 10^{-2}$  & $2.814\times 10^{-2}$  & $2.798\times 10^{-2}$ &  $2.860 \times 10^{-2}$ &  $2.673 \times 10^{-2}$  &  $2.650 \times 10^{-2}$ \\
		\bf ${\rm Rlff_D}$ &$ 0.6981$ & $0.6947$ & $0.8684$ & $0.6887$  & $0.7552$  & $0.6948$   \\
		\bf $M_{\rm A}$ & $8.121 \times 10^{-3} $& $7.868 \times 10^{-3} $ & $8.082\times 10^{-3}$ &$  7.866\times 10^{-3}$  &$  6.770\times 10^{-3}$  &$  6.789\times 10^{-3}$ \\
		\bf $R_{\rm A}$ & $0.3706$  & $0.3714$ & $0.3739$ &  $0.3705$   &  $0.3561$  &  $0.3555$ \\
		\bf $\rho_{\rm max,A}$ & $1.000$ & $1.000$& $1.000$ & $1.000 $  & $0.9580 $ & $0.9580 $ \\
		\bf $(M_{\rm c}/M)_{\rm A}$ & $9.001 \times 10^{-2} $& $9.850 \times 10^{-2}  $ & $9.913 \times 10^{-2} $ & $9.843 \times 10^{-2} $ & $9.589 \times 10^{-2} $ & $9.684 \times 10^{-2} $\\
		\bf ${\rm J_A}$ & $2.075\times 10^{-5} $& $1.945 \times 10^{-5} $ & $2.050\times 10^{-5}$ &$ 2.015 \times 10^{-5}$  &$1.380 \times 10^{-5}$  &$ 1.389 \times 10^{-5}$ \\
		\bf $K_{\rm c,A}$ & $8.418\times 10^{-3}$ & $8.310\times 10^{-3}$ &  $8.538\times 10^{-3}$ &  $ 8.320\times 10^{-3} $  &  $ 7.519\times 10^{-3} $  &  $ 7.527\times 10^{-3} $  \\
		\bf $K_{\rm e,A}$ & $3.384\times 10^{-2} $ & $3.381\times 10^{-2}$ & $3.473\times 10^{-2}$ & $ 3.384\times 10^{-2}$  & $ 3.087\times 10^{-2}$  & $ 3.090\times 10^{-2}$ \\
		\bf ${\rm Rlff_A}$ & $0.5146$  & $0.5079$ & $0.6394$ & $ 0.4941$  & $ 0.4293$   & $ 0.3963$ \\
		\bf $a$ & $1.134 $ & $1.141$ & $1.140$ &  $1.141	$  &  $1.140	$ &  $1.138	$ \\
		\bf $P$ & $64.15 $& $66.00$ & $65.51 $& $ 66.15$  & $ 73.48$  & $ 73.24$  \\
		\bf ${\rm J_{orb}}$ & $4.229 \times 10^{-4} $& $4.010 \times 10^{-4} $ &$4.068 \times 10^{-4}$ &$3.988 \times 10^{-4}$  &$2.816 \times 10^{-4}$  &$ 2.799 \times 10^{-4}$ \\
		\bf ${\rm VE}$ & $1.729 \times 10^{-3}$ & $4.968 \times 10^{-4}$ & $2.081 \times 10^{-4}$ & $2.698 \times 10^{-4}$  & $2.124 \times 10^{-5}$ & $-1.020 \times 10^{-4}$ \\
		\bf $\rho_{\rm min }$ &  $ 1.000 \times 10^{-10}$ & $ 1.000 \times 10^{-10} $  & $1.000 \times 10^{-10}$ & $1.000 \times  10^{-10}$  & $1.000 \times  10^{-10}$  & $1.000 \times  10^{-10}$\\		
                 \hline
	\end{tabular}
         \end{center}	
	\begin{tablenotes}
               \item {\ding{192}} The subscripts D, A, c and e stand for donor, accretor, core and envelope respectively.
         \end{tablenotes}	
\end{table*}

\begin{table*}
	\begin{center}
	\caption{Initial bipolytropic binary system parameters in solar units}
		\label{table:initbibi}
		\vspace{0.2cm}
	\begin{tabular}{lllllllllllll} 
		\hline
		 Model ID &  cylLR  &  cylHR &   octLR &  octHR  &   {cylEE} &   {octEE}   \\
		\hline
		 \bf $M_{\rm tot}$  & $1.702$ & $1.697$ & $1.676$ & $ 1.694$ & $ 1.598$ & $ 1.591$\\
		 \bf $M_{\rm A}$ & $1.000 $& $1.000 $&$ 1.000$ & $1.000  $  & $1.000  $  & $1.000  $\\
		  \bf $R_{\rm A}$ & $1.000 $& $1.000 $& $1.000 $& $1.000  $  & $1.000  $ & $1.000  $\\		 		 
		  \bf $M_{\rm D}$ & $0.7022 $& $0.6971$ & $0.6765$ &  $0.6936 $  &  $0.5979 $   &  $0.5905 $ \\
		  \bf $R_{\rm D}$ & $0.9425 $& $0.9430$ & $0.9351 $& $0.9426$  & $0.9500$  & $0.9449$ \\
		\bf $ a$ & $3.060 $& $3.073 $&$ 3.079$ &  $3.082 $  &  $3.079 $&  $3.074$ \\
		\bf $P(\rm hour)$ & $11.34$ & $11.45 $&$ 11.41$ & $11.52$  &    12.99  & $12.95$  \\
                 \hline
	\end{tabular}
         \end{center}	
\end{table*}

Before analysing the results, let us consider the outcome that we may expect from analytical approximations {for the two sets of simulations, Q0.7 and Q0.6}. There are two critical mass ratios which determine the ultimate outcome of a {conservative} mass transfer event. The first critical mass ratio is $q_a = 1$ for the separation; below this value the separation increases on mass transfer and above this value it decreases. The next critical mass ratio, $q_{\rm crit}$, arises from the adiabatic response of the donor as well as the response of its Roche lobe to mass loss. 
{This interplay is quantified by comparing the mass-radius exponent, $\zeta = {\rm d \hspace{0.05cm} log }R/ {\rm d\hspace{0.05cm} log} M$, of the Roche lobe ($\zeta_{\rm L}$) to that of the adiabatic value for the Roche lobe filling star ($\zeta_{\rm ad}$) \citep{Webbink-zeta1985}. Dynamically unstable mass transfer occurs when $\zeta_{\rm L} > \zeta_{\rm ad}$, and it continues in a quasi-steady state, on a thermal or nuclear timescale otherwise. For a conservative mass transfer, $\zeta_{\rm L} \approx 2.13q - 1.67$ \citep{Taut1997}.
The donor in the benchmark simulations has a deep convective envelope with polytropic index, $n_{\rm e}=3/2$. 
Thus for our analysis, we can assume an approximate value of $\zeta_{\rm ad} = -1/3$, corresponding to pure $n=3/2$ polytrope. This leads to the critical mass ratio, $q_{\rm crit} \approx 0.63$. 
For the benchmark simulations Q0.7 and Q0.6, the $\zeta_{\rm L}$ equals about -0.18 and -0.39 respectively. Thus mass transfer in Q0.7 simulation should be dynamically unstable, while that in the Q0.6 should be stable.
Note that simplifying assumptions are made in the above analytical treatment, such as point mass potentials and constant mass ratio during mass transfer. The behaviour of a real system depends on fully three dimensional calculations.
The initial mass ratio of 0.7 offers an interesting mass transfer case, as we can test behaviour of the simulations through the merger. The system with an initial mass ratio of 0.6 allows us to test if our tools are able to cope with simulations which undergo stable mass transfer.
We shall see in the next section that both the results of the benchmark simulations are consistent with the above expectations. 
Conversely, consistent hydrodynamical evolution confirms that the initial bipolytropic binary models obtained through the BSCF method are of sufficient quality to be useful as quiet initial conditions for such investigations. 
}

\section{Mass Transfer Simulations}
\label{mts}

{
In this section we describe the Q0.7 and Q0.6 set of simulations, as well as the framework for analysing the results.
Since the initial conditions were chosen to match closely, for the given mass ratio, we can expect agreement in the dynamical behaviour of these systems as well as the ultimate outcome. 
As explained in section \ref{numal}, the code Flow-ER evolves adiabatically with an ideal gas equation of state, while Octo-tiger takes into account shock-heating and uses the dual energy formalism with an ideal gas equation of state. 
The flow dynamics of the binary systems during mass transfer are not expected to be affected by these differences, as we shall also confirm in sections \ref{sec:hydro} and \ref{sec:hydro1}.

The BSCF method produces binary models which are detached in general.
A driving mechanism was incorporated in each hydrodynamic code such that a certain fraction of specific angular momentum was artificially removed from the system per orbit. 
This in turn reduced the separation thus eventually initiating mass transfer.
The lower mass star initially filled a larger fraction of its Roche lobe, as compared to its companion, in all of our models. Hence the mass transfer proceeded from the lower to the higher mass component.
The Q0.7 set of simulations was driven for 4 initial orbits by removing 2\% of the angular momentum per orbit, while the Q0.6 set was driven for 3 orbits at the rate of 1.5\% of the angular momentum per orbit. 
Note that as a consequence of the driving, the binary components were rotating non-synchronously at the onset of mass transfer. 
According to our experience of  studying binaries with polytropic components, ({$ \rm e.g.$} \cite{Motl2002}, \cite{Marcello2012}, \cite{Motl2017}) the driving phase does not affect the simulation results.
The subsequent hydrodynamical evolution of the binary systems was treated without any special assumptions or symmetries.}

\subsection{Diagnostics}
\label{sec:diag}
In order to study the evolution of the binaries in the {hydrodynamic} simulations, we analysed the temporal evolution of the following global parameters, called diagnostic quantities: 
\begin{enumerate} 
\item Binary separation, $a$
\item Mass transfer rate normalized to the initial mass and period, $ {|\dot{M}|}/{M_{\rm 0, ref}}=|{ {\rm d}M/ {\rm d}t}|/({M_{\rm 0}/P_{\rm 0}})$, for the donor 
\item Mass ratio, $ q =  M_{\rm D}/M_{\rm A}$ and normalized mass ratio, $ { q/q_{\rm 0}}$
\item Normalized orbital angular momentum, $ J_{\rm orb}/J_{\rm orb,0}$
\item Normalized spin angular momentum of the donor and accretor, $ J_{\rm D}/J_{\rm D,0}$, $ J_{\rm A}/J_{\rm A,0}$ 
\item Maximum density (also called central density) of the donor and accretor, $\rm \rho_{max, D}$ and $\rm \rho_{max, A}$ 
\end{enumerate}
Here a subscript ``0" denotes the initial value, ``D" denotes the donor and ``A" denotes the accretor. 
The diagnostic parameters were calculated in a similar way for the simulations conducted using both hydrodynamic codes.
The donor and accretor material were separated by a dividing plane, normal to the line joining the maximum density of the two stars at the ${\rm L_1}$ point.
{During the evolution, the ${\rm L_1}$ point was considered to be the point of the maximum effective gravitational potential along the line between the two centres of mass.}
The separation was calculated as the distance between the centre of mass of the donor and the accretor. The centre of mass, as well as the mass for each component, was calculated for the total mass on the either side of the dividing plane.
 The spin angular momentum of each component ($ J_{\rm D}$ and $J_{\rm A}$) was calculated in the inertial frame with respect to its centre of mass.
 The orbital angular momentum was calculated by subtracting the spin angular momentum of each component from the total angular momentum of the system, all calculated in the inertial frame.
 In addition to the above diagnostic quantities, we studied the density distributions in the equatorial as well as meridional plane. A data file containing the density field, called a ``frame", was written 120 times per initial orbit in Flow-ER and 100 times per initial orbit in Octo-tiger.  

The time in our simulations is always normalized with respect to the initial orbital period for the respective model. {For Q0.7 simulations, we synchronized the simulations with respect to a single event during the evolution, the merger of the binary occurring at time $t_{\rm merge}$. 
}We defined $t_{\rm merge}$ for the octHR simulation such that the the accretor spin angular momentum achieves the maximum value during the merger phase. We then compare the frame corresponding to $t_{\rm merge}$ for the octHR to the other simulations, by eye. The time corresponding to this closest matching frame is identified as $t_{\rm merge}$ for that particular simulation (see Figure \ref{fig:tmerge}). 
Once $t_{\rm merge}$ has been identified for each simulation, we could {align} all the time-dependent plots for the Q0.7 simulations. The particular choice of the accretor spin angular momentum is discussed later in section \ref{sec:hydro}.
The choice of the temporal ``zero point" is arbitrary and we chose the beginning of cylHR simulation as this point, since it completed the largest number of orbits. We shifted the diagnostic plots for the rest of the simulations such that the $t_{\rm merge}$ for each of them coincides. This shifted time was given by
\begin{equation}
t_{\rm shift}=t+t_{\rm zpt},
\end{equation}   
where
\begin{equation}
 t_{\rm zpt}=t_{\rm merge, \hspace{0.1cm} cylHR} - t_{\rm merge},
\end{equation}  
for each simulation.
We terminated the diagnostic plots soon after the $t_{\rm merge}$.

{In the case of Q0.6 simulations the binary systems exhibit a stable mass transfer throughout the evolution. Thus we do not have a clear event for synchronization and the diagnostic quantities are simply plotted from the beginning of the two simulations.

}

\subsection{Hydrodynamic Evolution {of Q0.7 Simulations}}
\label{sec:hydro}
\begin{figure*}
\begin {center}
\includegraphics [width=14.5cm] {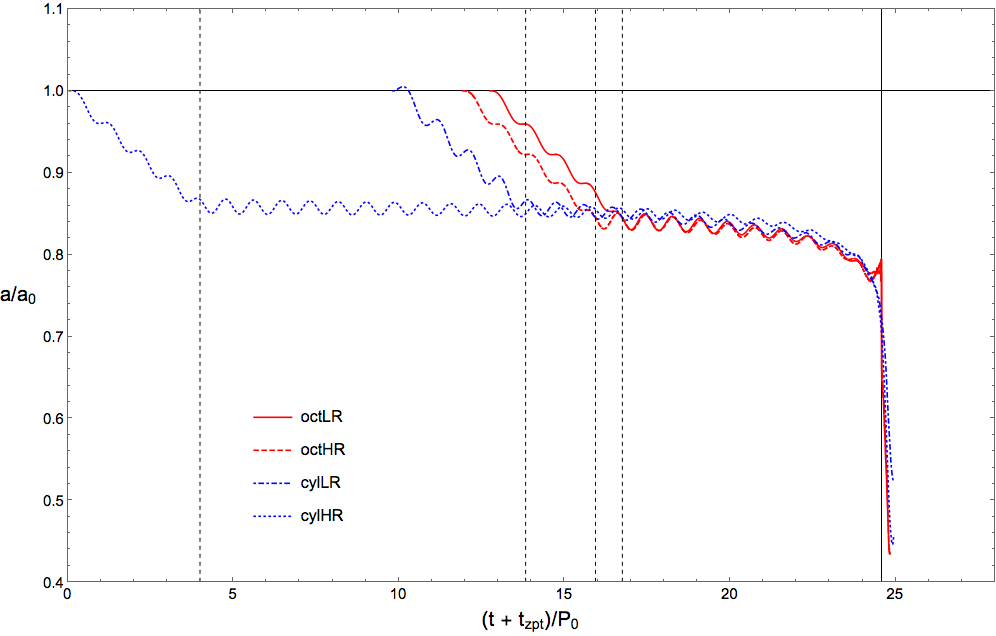}
\end {center}
\caption{Time-dependent behaviour of the normalized separation for Q0.7 simulations. 
The {vertical, dashed} lines mark the end of initial driving phase for cylHR, cylLR, octHR and octLR simulations respectively. The {vertical, solid} line marks $t_{\rm merge}$. The behaviour of the separation agrees quite well across all four simulations after the end of initial driving phase, $(t+t_{\rm zpt})/P_0 \approx 16.5$. The initial nearly linear decrease in the separation {was} caused by driving and the oscillations {were} caused by the slight {eccentricity} of the binary orbit due to perturbations from the driving.}
\label{fig:sep}
\end{figure*}

\begin{figure*}
	\begin {center}
	\includegraphics [width=15.5cm] {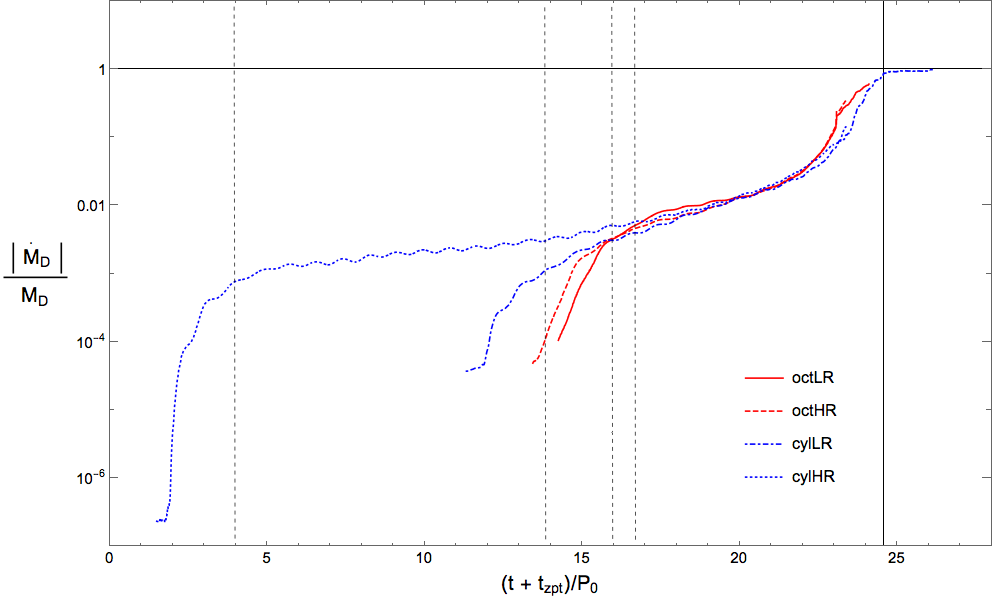}
	\end {center}
	\caption{Time-dependent behaviour of the smoothed mass loss rate of the donor for Q0.7 simulations. 
	The {vertical, dashed} lines mark the end of initial driving phase for cylHR, cylLR, octHR and octLR simulations respectively. The {vertical, solid} line marks $t_{\rm merge}$. The $ {|\dot{M}|}/{M_{\rm 0, ref}}$ plots agrees quiet well across all four simulations after the end of initial driving phase, $(t+t_{\rm zpt})/P_0 \approx 16.5$. The length of a simulation before the merger depends on the initial mass transfer rate. The difference in $ {|\dot{M}|}/{M_{\rm 0, ref}}$ between the two codes right before the merger {was} due to the difference in the treatment of shocks.
	}
	\label{fig:mdotsmoothd}
	\end{figure*}

\begin{figure*}
	\begin {center}
	\includegraphics [width=14.5cm] {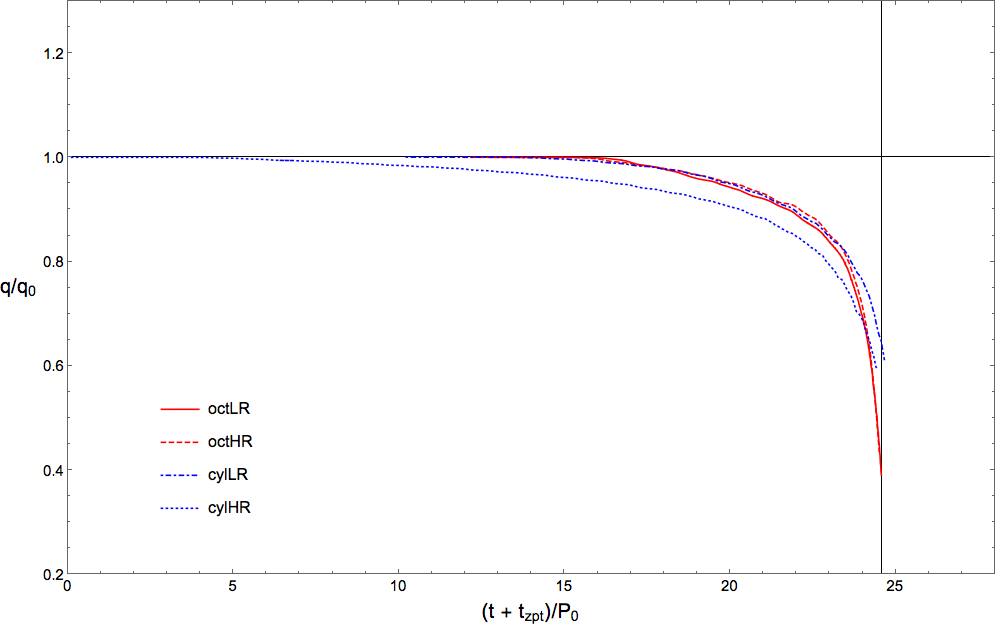}
	\end {center}
	\caption{Time-dependent behaviour of the normalized mass ratio for Q0.7 simulations. 
	The {vertical, solid} line marks $t_{\rm merge}$. 
	The mass ratio of cylHR {was} consistently lower because this simulation lasted almost twice as long when compared to the other three simulations.
	 }
	\label{fig:qnorm}
	\end{figure*}

\begin{figure*}
\begin {center}
\includegraphics [width=14.5cm] {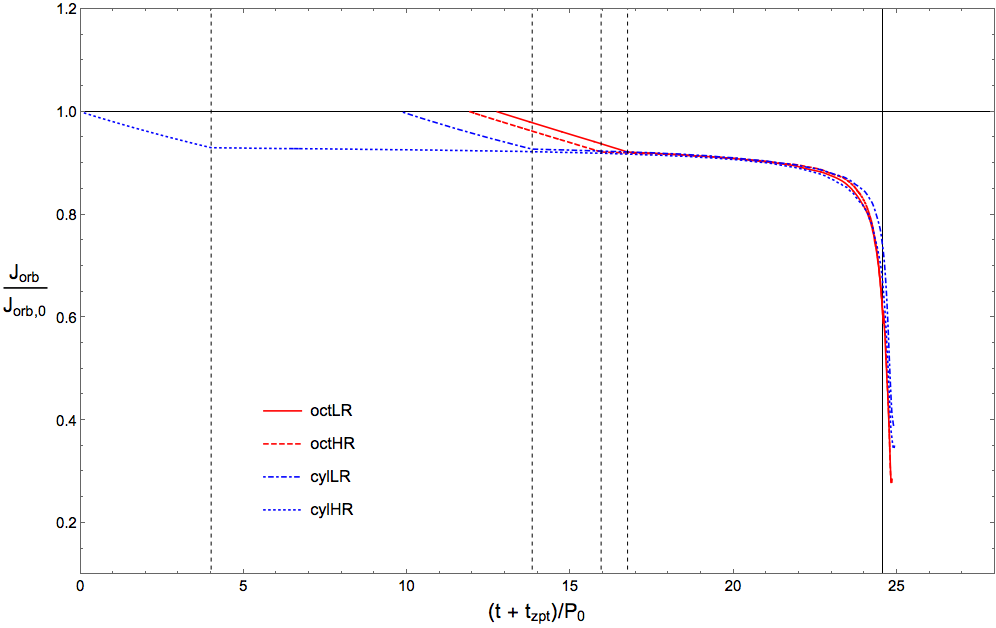}
\end {center}
\caption{Time-dependent behaviour of the normalized orbital angular momentum for Q0.7 simulations. 
{The vertical, dashed lines mark the end of initial driving phase for cylHR, cylLR, octHR and octLR simulations respectively. The vertical, solid line marks $t_{\rm merge}$.}
The linear drop in the beginning of each simulation was caused by the initial driving phase. 
}
\label{fig:jorb}
\end{figure*}

\begin{figure*}
\begin {center}
\includegraphics [width=14.5cm] {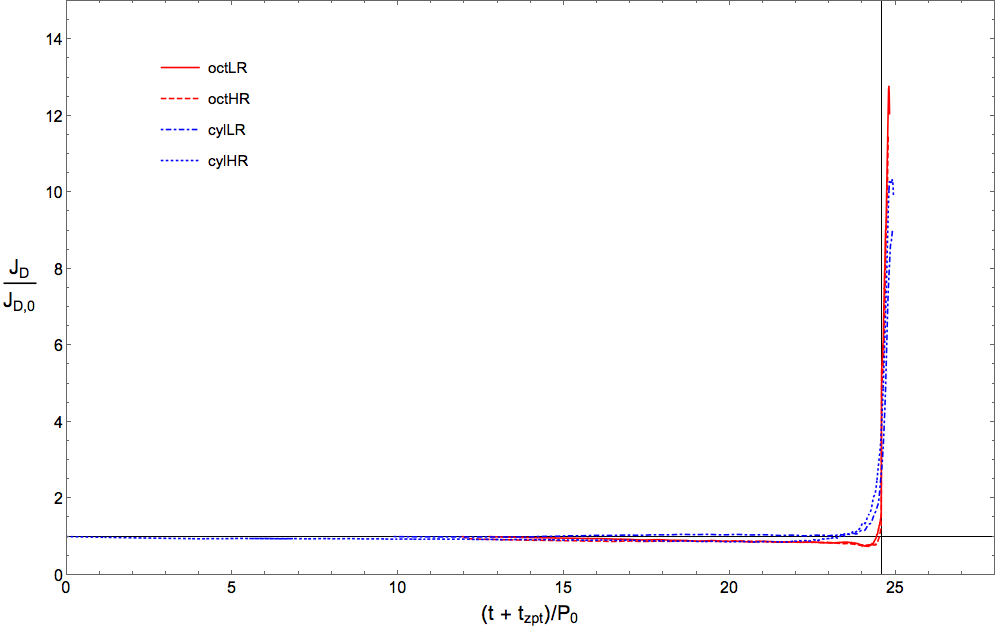}
\end {center}
\caption{Time-dependent behaviour of the normalized donor angular momentum for Q0.7 simulations. 
The {vertical, solid} line marks $t_{\rm merge}$. 
The difference in ${ J_{\rm D}/J_{\rm D,0}}$ between the two codes right before the merger {was} a reflection of different mass transfer rates.
}
\label{fig:jd}
\end{figure*}

\begin{figure*}
\begin {center}
\includegraphics [width=14.5cm] {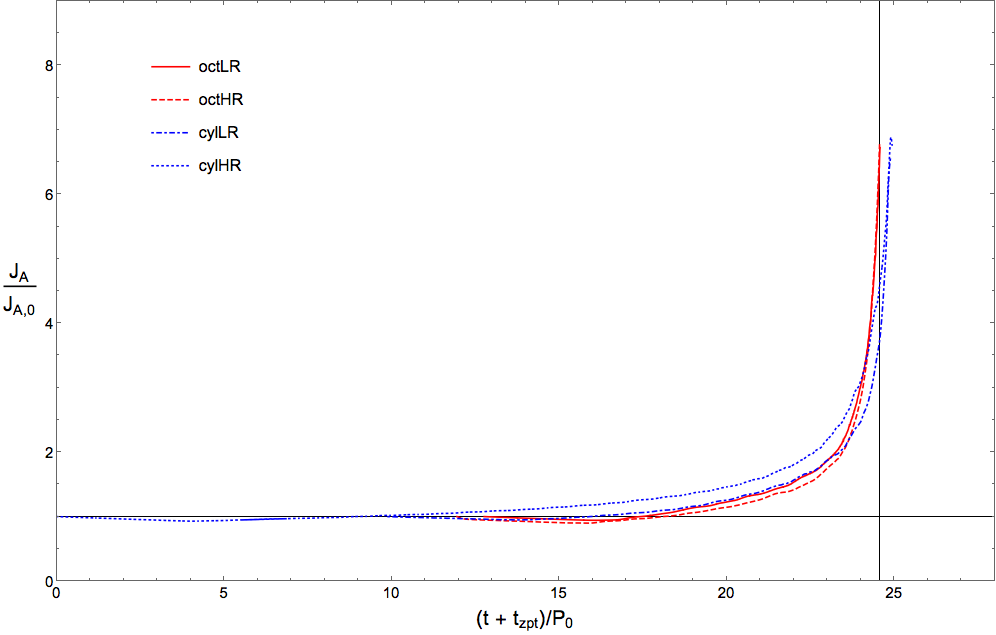}
\end {center}
\caption{Time-dependent behaviour of the normalized accretor angular momentum for Q0.7 simulations. 
The {vertical, solid} line marks $t_{\rm merge}$. 
The accretor angular momentum of cylHR {was} consistently higher because this simulation lasted almost twice as long when compared to the other three simulations.
}
\label{fig:ja}
\end{figure*}

\begin{figure*}
\begin {center}
\includegraphics [width=14.5cm] {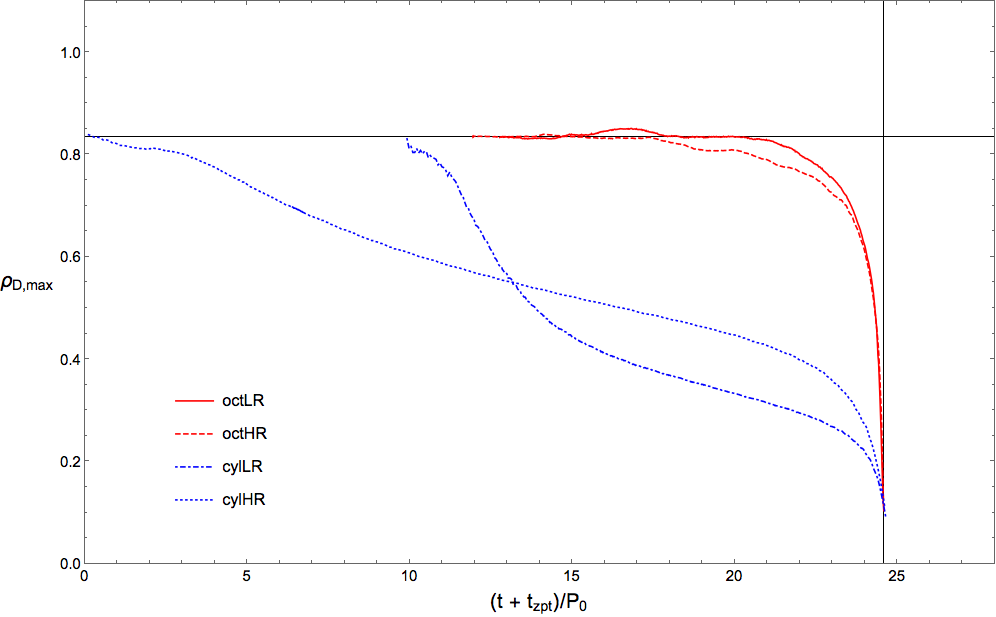}
\end {center}
\caption{Time-dependent behaviour of the donor maximum density for Q0.7 simulations. 
The vertical line marks $t_{\rm merge}$.
The horizontal line marks the approximate value of the initial donor maximum density. The fall in the maximum density in the case of Flow-ER simulations {was} a numerical artefact caused by poorly resolved core.}
\label{fig:rhod}
\end{figure*}

\begin{figure*}
\begin {center}
\includegraphics [width=14.5cm] {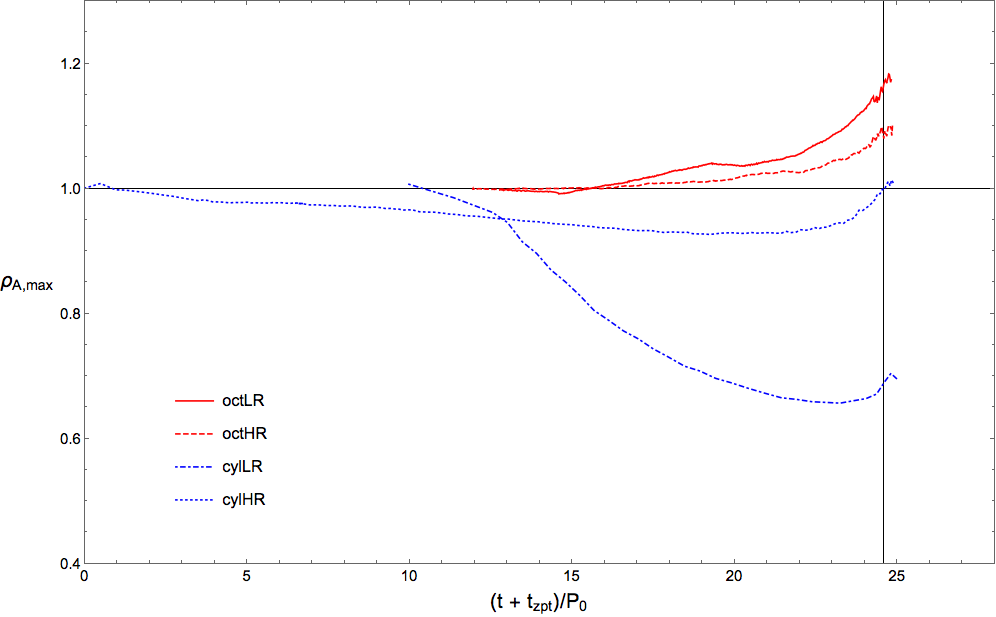}
\end {center}
\caption{Time-dependent behaviour of the accretor maximum density for Q0.7 simulations. 
The vertical line marks $t_{\rm merge}$.
The horizontal line marks the approximate value of the initial accretor maximum density. The fall in the maximum density in the case of Flow-ER simulations {was} a numerical artefact caused by poorly resolved core.}
\label{fig:rhoa}
\end{figure*}

{
In this section we describe the evolution of Q0.7 set of binaries and discuss the results.  
As explained in section \ref{tim}, the initial mass ratio lies above the critical value for the system, $q_{\rm crit} \approx 0.63$, hence we expect an unstable mass transfer over a dynamical timescale. }
The benchmark for these simulations was to evolve the initial bipolytropic binary to a fixed point in time, ${\rm i.e.}$ to merger.
Table \ref{table:simsum} describes the summary of the Q0.7 simulations along with some quantities of interest, and Table \ref{table:movies} lists the movies of the temporal evolution of the equatorial cross sections of the density fields.
The time-dependent behaviour of key diagnostic parameters are plotted in Figures \ref{fig:sep}-\ref{fig:ja}, and Figures \ref{fig:rhod} and \ref{fig:rhoa}.
We found that all four binary models were unstable to mass transfer, resulting ultimately in a merger. 
The end states as well as the intermediate behaviour agree very well with the simple analytical approximations and also among the simulations.
Although there were a few key differences among the simulations, all of these can be accounted for by our current understanding of the numerical tools.

\begin{table*}
	\caption{ Quantities of interest for Q0.7 simulations}
		\label{table:simsum}
	\vspace{0.2cm}
	\begin{center}
	\begin{tabular}{lllllllllllll} 
		\hline
		 Model ID &  cylLR  &  cylHR &   octLR &  octHR    \\
		\hline
		{Equation of state {\ding{192}}  }  &  Ideal gas & Ideal gas & Ideal gas with  & Ideal gas with    \\
		  		        &   & & Dual Energy & Dual Energy  \\		  
		                        &   & & Formalism & Formalism  \\			 
{Total orbits ($ t/ \rm P_0$)}   & $17.77$  & $25.20$ & $14.89 $& $ 13.59$  \\
		{${ t_{\rm merge} (t/P_0)} $ \ding{193} } & $15.02$  & $24.58$ & $12.10 $& $ 12.91$  \\
		   {${ t_{\rm zpt}  (t/P_0)} $ \ding{194} }& $9.83$  & $0$ &$ 12.75$ &  $11.94  $\\
		  Timesteps  &  $ 1.84$ M & $ 7.02$ M & $  160$ k & $ 525$ k   \\
		 {${\rm T_{\rm wall} (d)}$ \ding{195} } & $6 $ & $177 $& $16 $ & $ 38 $ \\
		{Number of cores}  & $256$  & $256$ &$ 1280$ & $ 4000$  \\		  
		 {Cost (CPU-hr/orbit) }            &  $ 2.07 $ k & $ 43.2$ k & $ 33.0$ k & $ 268$ k   \\
		 {Cost (calculations/orbit) }     &  $ 1.74 \times 10^{16}  $  &  $ 3.62 \times 10^{17}  $ &  $ 2.77 \times 10^{17}  $&  $ 2.25 \times 10^{18}  $  \\
		  Machine  &  QueenBee   & QueenBee & SuperMIC  & SuperMIC    \\
		    &   (LONI)&(LONI)  & (LSU HPC) &  (LSU HPC)  \\
		  Processors  & 2.33GHz 4-Core   & 2.33GHz 4-Core & 2.8GHz 10-Core  &  2.8GHz 10-Core  \\
		    & Xeon 64-bit & Xeon 64-bit & Ivy Bridge-EP  & Ivy Bridge-EP \\
		    & & &  Xeon 64-bit & Xeon 64-bit\\
                 \hline
	\end{tabular}
	\end{center}
	\begin{tablenotes}
               \item 
            {\ding{192}} Equation of state used in the simulation for determining the pressure.
            {\ding{193} }The time of merger (as determined in section \ref{sec:diag}).
             {\ding{194} }The difference between ${\rm t_{merge}} $ for cylHR simulation and the current one (Equation 6.2).
             {\ding{195}} The time needed to complete the simulation. 
         \end{tablenotes}
         \vspace{-0.1cm}
\end{table*}

\begin{table*}
	\begin{center}
	\caption{Simulation movies}
		\label{table:movies}
		\vspace{0.2cm}
	\begin{tabular}{lllllllll} 
		\hline
		 Model ID &  {Movies \ding{192}}   \\
		\hline
		 cylLR &  cylLR   \\  
		cylHR  &  cylHR \\
		 octLR & {octLR-zoomed \ding{193}} &   {octLR-wide \ding{194}}   \\
		 octHR &   octHR-zoomed  &   octHR-wide   \\
                 \hline
	\end{tabular}
	\end{center}
		\begin{tablenotes}
               \item 
               {\ding{192} The movies are available for viewing in the online supplemental material.}
               {\ding{193}} Zoomed movies show equatorial cross sections at the length scale of the cylLR grid.
               {\ding{194}} Wide movies show the full octHR grid.
         \end{tablenotes}
\end{table*}

Consider the separation between the two components, plotted in Figure \ref{fig:sep}. Oscillations with an amplitude of approximately $2\%$ can be noticed throughout the evolution up to merger for all four simulations. In general, an initial binary model obtained through the BSCF method is not in perfect equilibrium and the orbits are not perfect circles. The Q0.7 binaries were further perturbed by the artificial removal of the angular momentum during the initial driving phase, exacerbating the situation. 
The slight deviations from equilibrium resulted in binary orbits with small {eccentricity}, in turn causing the oscillations of the components about the mean separation.
The final amplitude of the oscillations also depends on their exact phase when the driving is stopped, hence
the absolute phase of these oscillations should not be expected to match for any two simulations.
The next prominent feature is an initial, roughly linear fall in the separation for the first 4 orbits, which was a direct result of the artificial removal of angular momentum, as each binary was driven together. 
The logarithmic derivative of the separation, $\dot{a}/a$, is proportional to twice that of the angular momentum, $\dot{J}/J$ for a conservative mass transfer \citep{APIA}. A total of $8\%$ decrease in the total angular momentum thus resulted in about $16\%$ decrease in the separation, as observed in Figure \ref{fig:sep}.
We made sure that a steady mass transfer stream was established before the driving was stopped.
The behaviour of the Q0.7 systems agreed remarkably well after this point, ${\rm i.e.}$ after $(t+t_{\rm zpt})/P_0 \approx 16.5$. 
The separation decreased gradually over about six orbits, until the binary merges.
The merger was rapid and occurred within only about two orbits.

Figure \ref{fig:mdotsmoothd} shows the normalized and smoothed mass transfer rate of the donor ($ {|\dot{M}|}/{M_{\rm 0, ref}}$).
 Note that the ordinate is in a logarithmic scale. 
We needed to use the smoothed rates, since the plots for the instantaneous mass transfer rate turned out to be very noisy because of the subtraction of two nearly equal quantities both in the numerator as well as the denominator.
 The smoothing was performed using a moving boxcar average with a width of three initial orbital periods.
 The relatively low amplitude oscillations in the $ {|\dot{M}|}/{M_{\rm 0, ref}}$ before the sharp rise near the end are caused by the change in the mass transfer rate due to the epicyclic oscillations in the orbital separation.
Once the driving ended and a steady mass transfer began at $(t+t_{\rm zpt})/P_0 \approx 16.5$,  $ {|\dot{M}|}/{M_{\rm 0, ref}}$ increased steadily as the binaries evolved. The average rate of the normalized mass transfer after $(t+t_{\rm zpt})/P_0 \approx 16.5$ was of the order of 0.1 per orbit, which explains the lifetime of the binary with the merger occurring within about 8 orbits.
The cylHR simulation lasted the longest because at the end of its driving,  $ {|\dot{M}|}/{M_{\rm 0, ref}} \approx 10^{-5}$, while octLR simulation lasted the shortest time as the mass transfer rate at the end of its driving phase was the highest amongst the Q0.7 simulations, $ {|\dot{M}|}/{M_{\rm 0, ref}} \approx 5 \times 10^{-3}$. 
The shifted plots for  $ {|\dot{M}|}/{M_{\rm 0, ref}}$ matched very well for all simulations after the end of the initial driving phase at $(t+t_{\rm zpt})/P_0 \approx 16.5$, until about two orbits before the merger.

{As shown in Table \ref{table:simsum},} the time taken from the beginning of the simulation to the merger, $t_{\rm merge}$, varied by a large factor across the simulations.
For the Flow-ER simulations, $t_{\rm merge}$ was longer as compared to the Octo-tiger simulations, and it was also longer at the higher resolutions for both the codes.
We can explain the observed trend in $t_{\rm merge}$ as follows. 
The time of merger in a simulation depends on the average mass transfer rate, which in turn depends on the depth of initial contact as well as the resolution. 
First consider the trend across the two codes, since they have a key difference -- Octo-tiger evolves the total energy equation explicitly which allows for shock heating of the gas.
Thus in Octo-tiger simulations, when the accretion stream impacts the accretor, the supersonic velocities shock-heat the outer atmosphere of the stars, forming a puffed up common envelope-like structure.
This condition is exacerbated by the fact that radiative transport is not implemented in the code, which prevents the material from cooling. 
With a hot atmosphere, the mass transfer rate should be higher in Octo-tiger simulations, as it should scale with the sound speed of the fluid near the ${\rm L_1}$ point.
This common envelope can also possibly cause numerical drag on the binary, leading to a relatively rapid merger even at the same level of initial contact depth. 
As seen in Figure \ref{fig:mdotsmoothd}, all simulations showed a sharp increase in $ {|\dot{M}|}/{M_{\rm 0, ref}}$ near the end, which indicates a rapid merger. 
However, during the final phase of the merger, there {was} a noticeable difference between the Flow-ER and the Octo-tiger simulations.
We conjecture that the higher mass transfer rate during the merger process in the Octo-tiger simulations was also caused by the hotter atmosphere of the donor. For a given code, the higher resolution simulations lasted longer because a finer grid can resolve a lower mass transfer rate.  
We will show that the trends in the diagnostic plots support this hypothesis.

	Figure \ref{fig:qnorm} shows the normalized mass ratio of the Q0.7 simulations. 
	The four curves in this plot essentially represent the history of mass transfer in each simulation.
	As the binaries evolved, $q$ decreased monotonically, and it continued to decrease even after driving was stopped, indicating a continuing mass transfer from the lower mass component to its companion. 
	The $q/q_0$ plots for all Q0.7 simulations, except for the cylHR, show a high degree of agreement.  
	The discrepancy between the evolution of the mass ratio of the cylHR simulation and the other simulations combined can be explained as follows.
	As a simulation progresses, if we assume the same mass transfer rate, the mass ratio depends on the initial masses of both the components in a binary as well as the amount of mass transferred from the donor to the accretor. 
	Consider Figure \ref{fig:mdotsmoothd} again, which shows that the mass transfer in the case of the cylHR simulation lasted almost twice as long as the other three simulations.
	Before we can expect an agreement between the codes, ${\rm i.e.}$ before $(t+t_{\rm zpt})/P_0 \approx 16.5$, the donor in the cylHR simulation was already transferring mass for about 12.5 orbits. Over this period, the normalized mass transfer rate can be considered as approximately $2 \times 10^{-3}$ per orbit. Thus the donor had transferred about 2.5 \% of its mass before $(t+t_{\rm zpt})/P_0 \approx 16.5$, making the mass ratio smaller by the same factor, as observed in Figure \ref{fig:qnorm}.
	Notice that over the last two orbits, the Octo-tiger simulations show a steeper slope of $q/q_0$ as compared to the Flow-ER simulations which indicates a larger rate of mass transfer, as we have already seen in Figure \ref{fig:mdotsmoothd}.

	The normalized orbital angular momenta are plotted in Figure \ref{fig:jorb}, for the Q0.7 simulations. Since all binaries were driven for four orbits at the rate of 2\% per orbit, the orbital angular momentum for each simulation initially showed a linear decline of a total of 8\%. 
	Figures \ref{fig:jd} and \ref{fig:ja} show the plots for the normalized spin angular momenta of the donor and accretor respectively. 
	First consider the qualitative behaviour of these plots during the mass transfer.
	For all Q0.7 simulations, ${ J_{\rm orb}/J_{\rm orb,0}}$ decreased steadily even after the driving {was} stopped and just before the merger, it fell precipitously over the last two orbits. We see that as ${ J_{\rm orb}}$ decreased right before the merger, both ${J_{\rm D}}$ and ${ J_{\rm A}}$ peaked sharply, indicating that the orbital angular momentum was being used to spin up each of the binary components. This behaviour of the angular momenta was a clear indication of tidal instability of the binary system. 
	With this instability, each component tried to maintain synchronicity, which extracted angular momentum from the orbit, resulting in a merger.
	In certain cases the binary can develop a mass-transfer instability, wherein the donor keeps transferring mass to the accretor in a dynamically unstable manner, such that the separation decreases and the donor essentially ends up ``falling" on the accretor. 
	Thus the spin angular momentum of the donor remains relatively constant until the merger, {unlike the results of} Q0.7 simulations. 
	We determined $t_{\rm merge}$ on the basis of ${ J_{\rm A}}$ {(section \ref{sec:diag})} because in all cases, the accretor angular momentum grows rapidly during the final stages of binary merger.

All three angular momenta (orbital and spin for both the components) generally agree across the Q0.7 simulations, however, there are two differences. The accretor angular momentum of the cylHR simulation starts to deviate early on from the corresponding curves of the other simulations, and the two codes behave differently during the merger over the last two orbits before $t_{\rm merge}$.
Consider the evolution of ${ J_{\rm D}}$ in the cylHR simulation. As seen earlier, the donor had transferred about 2.5\% of its mass before $(t+t_{\rm zpt})/P_0 \approx 16.5$. This mass also carried angular momentum, which resulted in spinning up of the accretor.
The component of the velocity of the accreting material at the ${\rm L_1}$ point, perpendicular to the line joining the two stars can be given as $v_{\perp} = R_{LA} \times \Omega$, where $R_{LA}$ is the Roche lobe radius of the accretor. 
Using the values in Table \ref{table:initbibicode}, we can estimate that the ${\approx 2.5 }$\% mass lost by the donor transferred about $4.6$\% of its angular momentum to the accretor before $(t+t_{\rm zpt})/P_0 \approx 16.5$. 
The accretor also had sufficient time to interact with, and gain angular momentum from, ${ J_{\rm orb}}$.
The combination of these two factors can explain the observed discrepancy of ${ J_{\rm A}}$ in Figure \ref{fig:ja}. 
During the final two orbits before the merger, the behaviour of the spin angular momentum of the donor as well as the accretor {was} different between the two codes. The sharper rise in the slope of both ${ J_{\rm D}}$ and ${ J_{\rm A}}$ in the case of the Octo-tiger simulations suggests that the orbital angular momentum was extracted more efficiently during the merger. This could be facilitated by a higher mass transfer rate near the end. Thus the difference in the spin angular momenta between the Flow-ER and Octo-tiger simulations is consistent with our analysis of the mass transfer rate earlier in this section.
The agreement between the diagnostic plots indicates that the initial Q0.7 binary models generated using the BSCF methods were sufficiently closely matched.

{
The spin up of the stars at the expense of the orbital angular momentum was rapid and occurred over a dynamical timescale. From the initial values of the angular momenta in Table \ref{table:initbibicode}, it is clear that the spin angular momenta remained much smaller than the orbital angular momentum until the merger. Thus the \cite{Darwin1880} instability can not play a role in the dynamical evolution.
The tidal synchronization timescale for a non-interacting binary with parameters matching the initial systems, {$\rm e.g.$} as estimated by \cite{Zahn1977}, is of the order of $10^5$ years. However, this is in the case of a detached binary system.
The binaries in Q0.7 simulations undergo unusually high rate of mass transfer. At the end of initial driving phase, the rate of mass transfer was about $10^{-3} M_{\odot}/$orbit  as compared to about $10^{-5} M_{\odot}/$yr for observed Algol-like systems. This resulted in a corresponding shortening in the response of the individual spins and a relatively rapid spin evolution. }

\begin{figure*}
  \includegraphics[width=17.6cm]{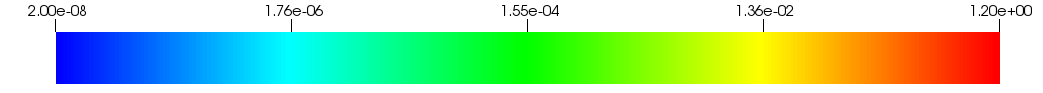}
  \vspace{0cm}
  \hspace{0cm}%
\stackunder[-2pt]{\includegraphics[width=8.6cm]{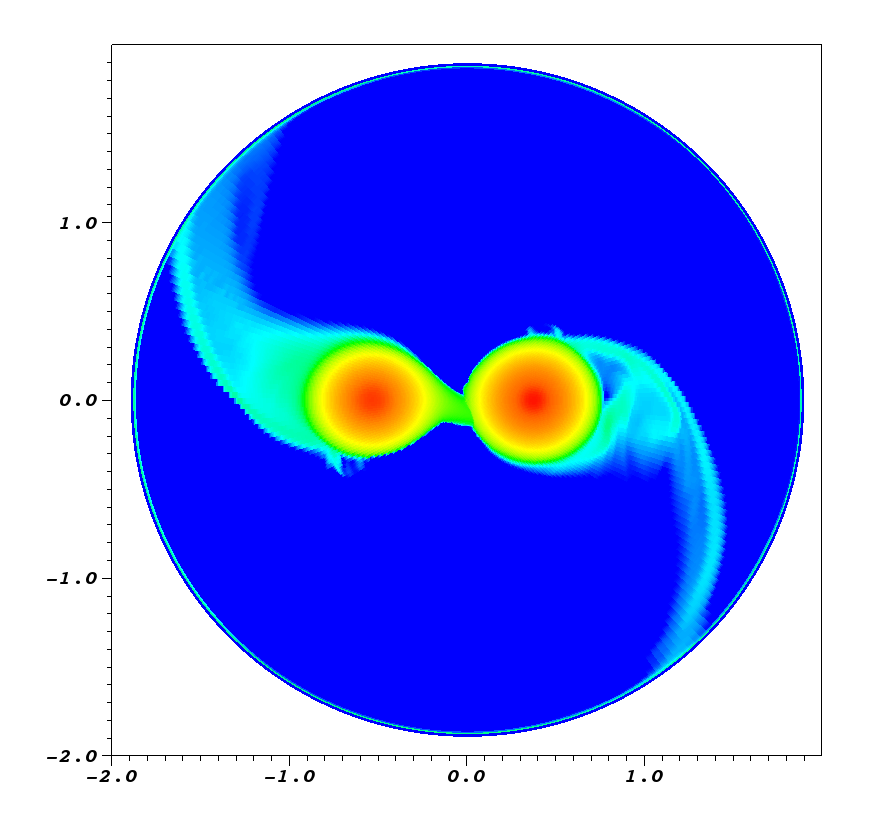}}{   (a) cylLR}%
  \hspace{0cm}%
\stackunder[-2pt]{\includegraphics[width=8.6cm]{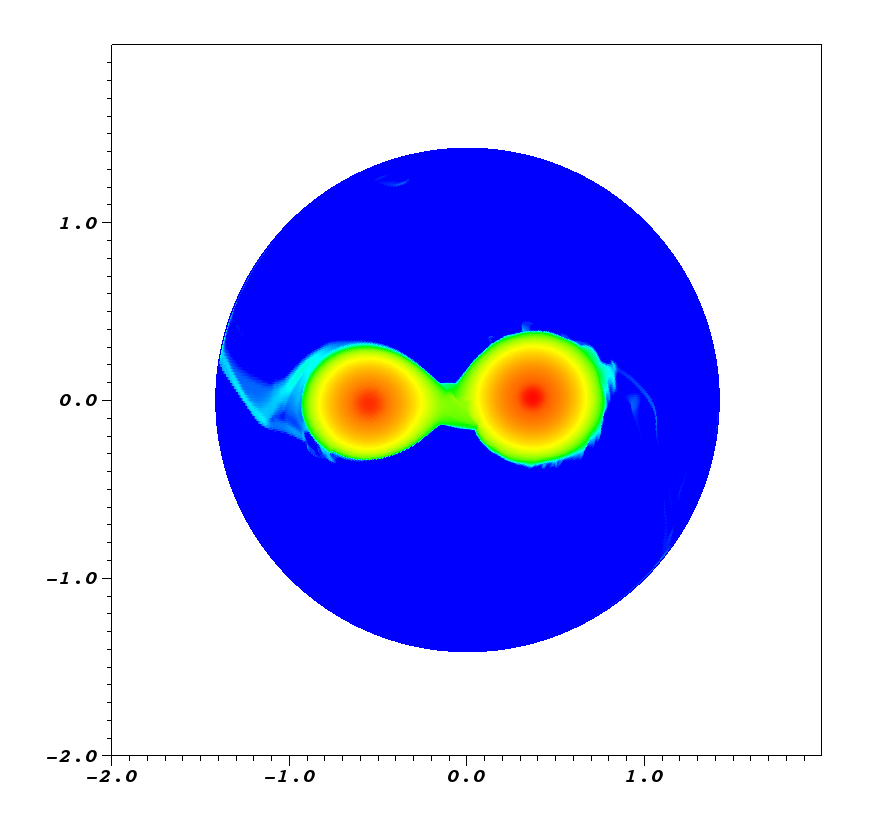}}{  (b) cylHR}
  \vspace{0cm}
  \hspace{0cm}%
\stackunder[-2pt]{\includegraphics[width=8.6cm]{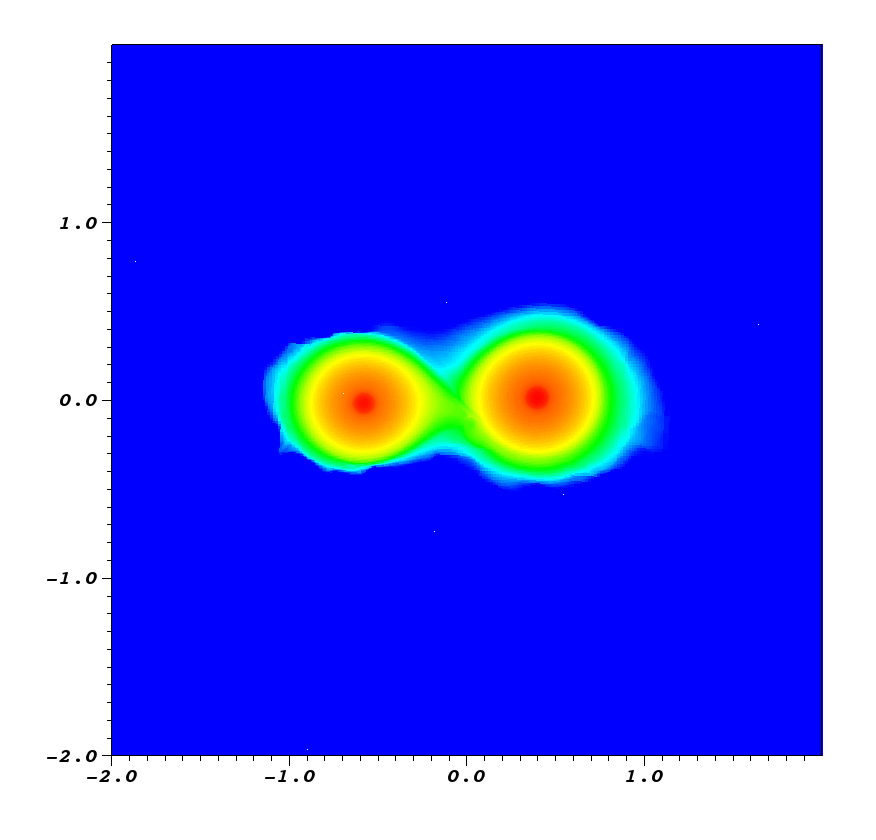}}{  (c) octLR}%
  \hspace{0cm}%
\stackunder[-2pt]{\includegraphics[width=8.6cm]{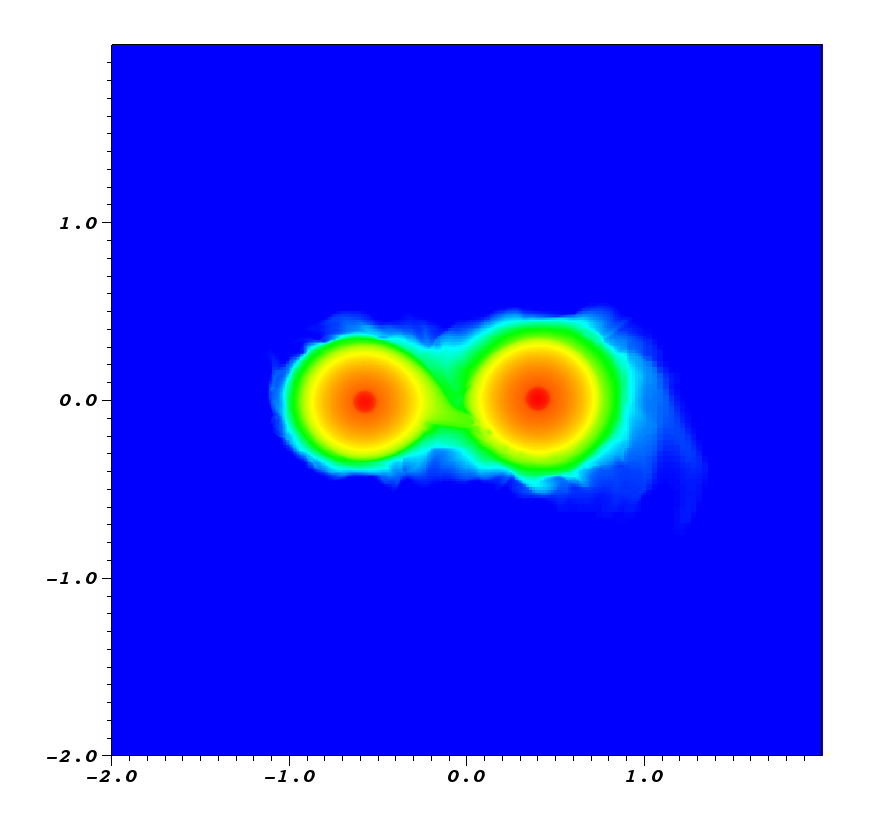}}{ (d) octHR}
\caption{Equatorial cross section of the density field at ${ \rm t_{merge} - 8.0}$ or $(t+t_{\rm zpt})/P_0= 16.58$. The diagnostic quantities of the simulations {agreed} after this point. This view also shows resolution dependent numerical wind in the Flow-ER simulations and the shock-heated atmosphere of the binary systems in the Octo-tiger simulations.}
\label{fig:tmerge-alot}
\end{figure*}

\begin{figure*}
  \includegraphics[width=17.6cm]{figs/color2.png}
  \vspace{0cm}
  \hspace{0cm}%
\stackunder[-2pt]{\includegraphics[width=8.6cm]{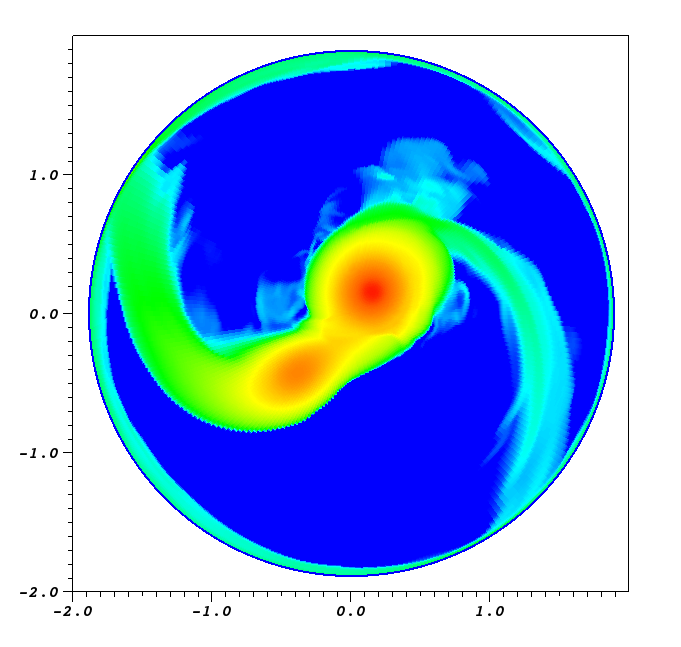}}{   (a) cylLR}%
  \hspace{0cm}%
\stackunder[-2pt]{\includegraphics[width=8.6cm]{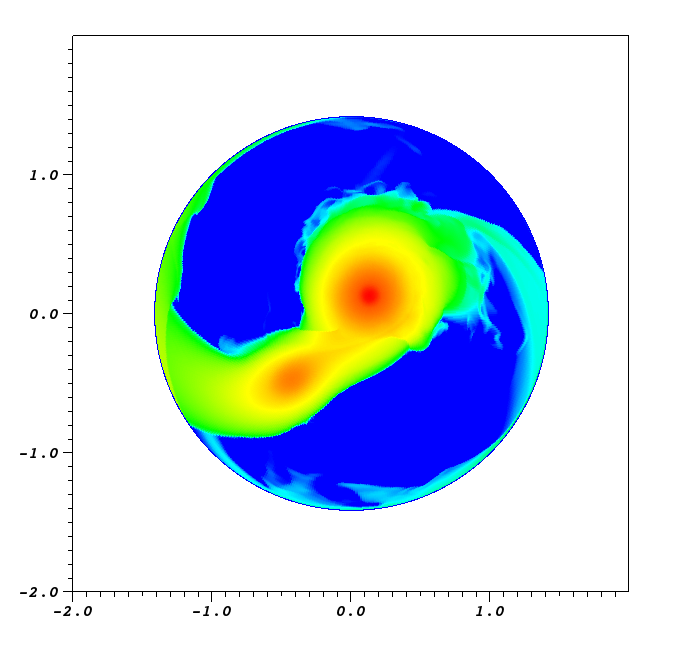}}{  (b) cylHR}
  \vspace{0cm}
  \hspace{0cm}%
\stackunder[-2pt]{\includegraphics[width=8.6cm]{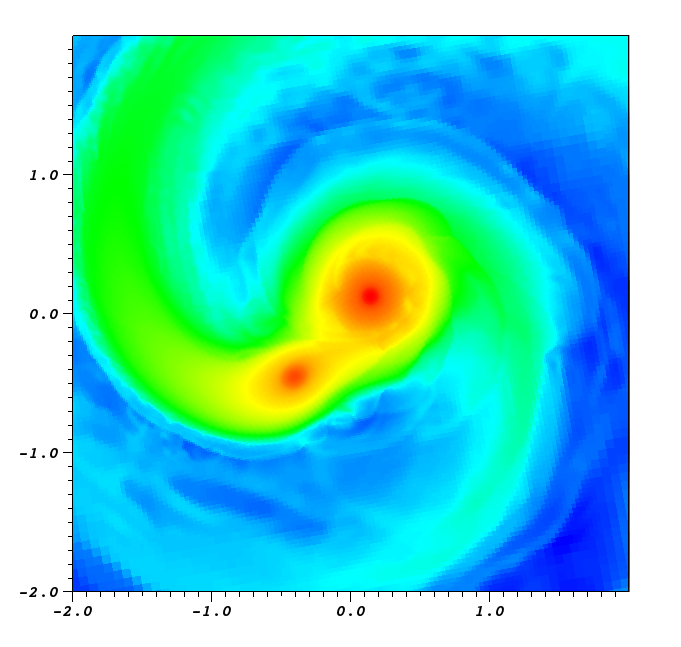}}{  (c) octLR}%
  \hspace{0cm}%
\stackunder[-2pt]{\includegraphics[width=8.6cm]{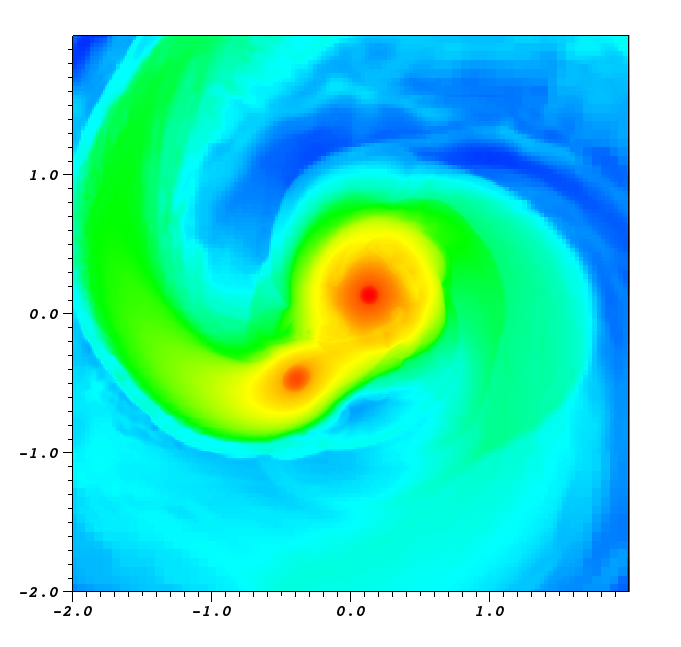}}{ (d) octHR}
\caption{Equatorial cross section of the density field at ${ \rm t_{merge}}$ occurring at $(t+t_{\rm zpt})/P_0= 24.58$, as determined in section \ref{sec:diag}.}
\label{fig:tmerge}
\end{figure*}

\begin{figure*}
  \includegraphics[width=17.6cm]{figs/color2.png}
  \vspace{0cm}
  \hspace{0cm}%
\stackunder[-2pt]{\includegraphics[width=8.6cm]{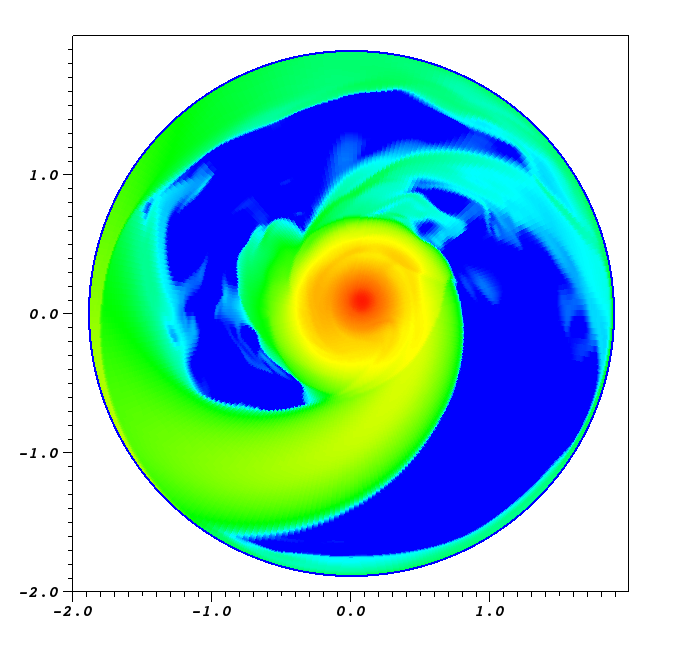}}{   (a) cylLR}%
  \hspace{0cm}%
\stackunder[-2pt]{\includegraphics[width=8.6cm]{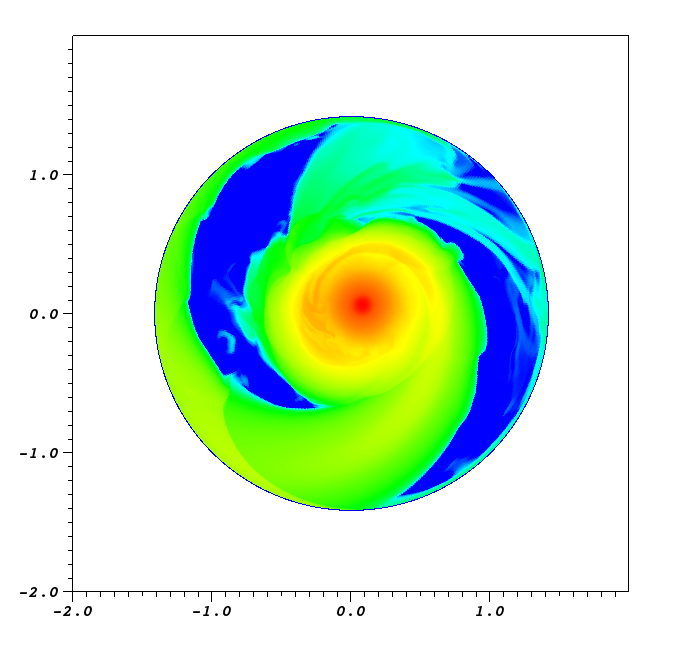}}{  (b) cylHR}
  \vspace{0cm}
  \hspace{0cm}%
\stackunder[-2pt]{\includegraphics[width=8.6cm]{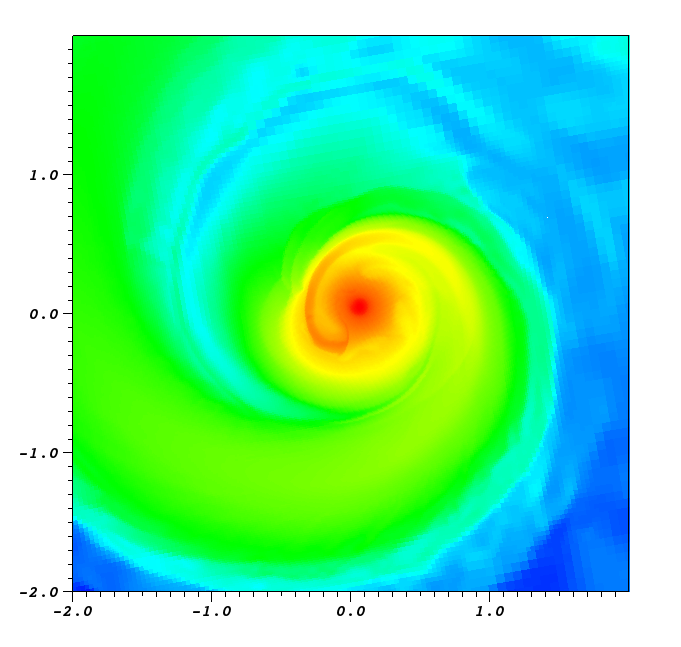}}{  (c) octLR}%
  \hspace{0cm}%
\stackunder[-2pt]{\includegraphics[width=8.6cm]{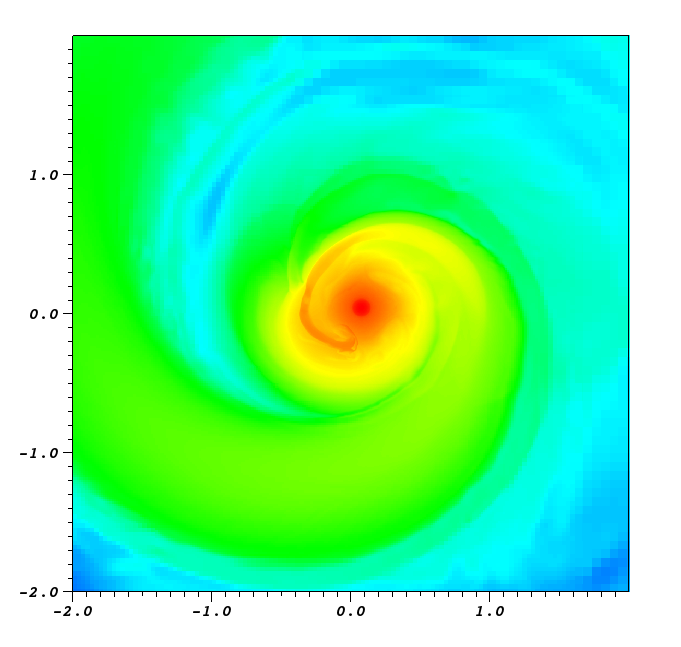}}{ (d) octHR}
\caption{Equatorial cross section of the density field at ${ \rm t_{merge} + 0.4}$ or $(t+t_{\rm zpt})/P_0= 24.98$, right after the merger. The disrupted core of the donor can be seen getting wrapped around that of the accretor.}
\label{fig:tmerge+0p4}
\end{figure*}

Figures \ref{fig:tmerge-alot}-\ref{fig:tmerge+0p4} show equatorial cross sections of the density fields during the evolution of the Q0.7 simulations. 
We used an open source visualization tool VisIt \citep{visit} to render the cross sectional frames for the simulations.
The distance scale used {was} the same for each simulation and each frame {was} 4 code length-units wide.
The density field ranged from $2 \times 10^{-8}$ to $1.2$ in code units and is plotted on a logarithmic scale, with the intention of visualizing the low density accretion stream, while also depicting the core-envelope structure. 
Each figure corresponds to the same $(t+t_{\rm zpt})/P_0$ during the binary simulation, thus representing the same stage of the dynamical evolution.
Each figure is also rotated through a trivial angular phase about the axis of rotation to match the octHR simulation, so that the key features can be directly compared.
 Figure \ref{fig:tmerge-alot} corresponds to $(t+t_{\rm zpt})/P_0= 16.58$ when a steady, direct impact mass transfer stream was established. 
The time is close to the end of the driving phase of the octLR simulation (see Figure \ref{fig:jorb}). The behaviour of the Q0.7 systems agreed remarkably well after this point, with the exceptions discussed above. 
The resolution-dependent ``numerical wind" can be noticed in the Flow-ER simulations. This figure also shows the formation of a puffed up, common envelope in the Octo-tiger simulations, due to the shock-heated atmosphere. Note that a real outflow of material may occur in binaries from these outer Lagrangian points during mass transfer.

Figure \ref{fig:tmerge} was used to determine $t_{\rm merge}$ for all simulations by matching as closely as possible with the octHR density. 
All four simulations showed the inspiral of the Q0.7 binary systems during the merger and the material lost through the outer Lagrange points, $\rm L_2$ and $\rm L_3$, forming a double pinwheel pattern. 
The contrast between the core and the envelope in the Flow-ER simulations was noticeably less sharp as compared to the Octo-tiger ones, especially in case of the donor star.
This is because in the Flow-ER simulations, the contact discontinuity at the core-envelope boundary dissolved with time due to insufficient resolution.

\begin{figure*}
  \includegraphics[width=17.6cm]{figs/color2.png}
  \vspace{0cm}
  \hspace{0cm}%
\stackunder[-2pt]{\includegraphics[width=8.6cm]{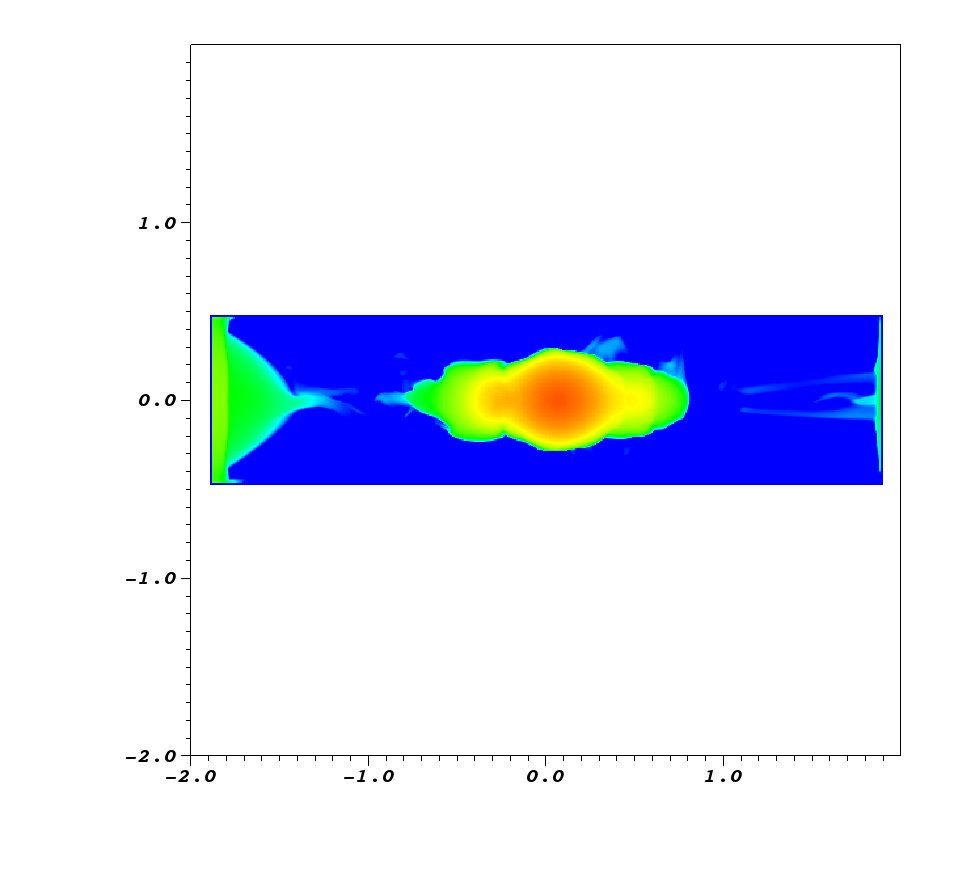}}{   (a) cylLR}%
  \hspace{0cm}%
\stackunder[-2pt]{\includegraphics[width=8.6cm]{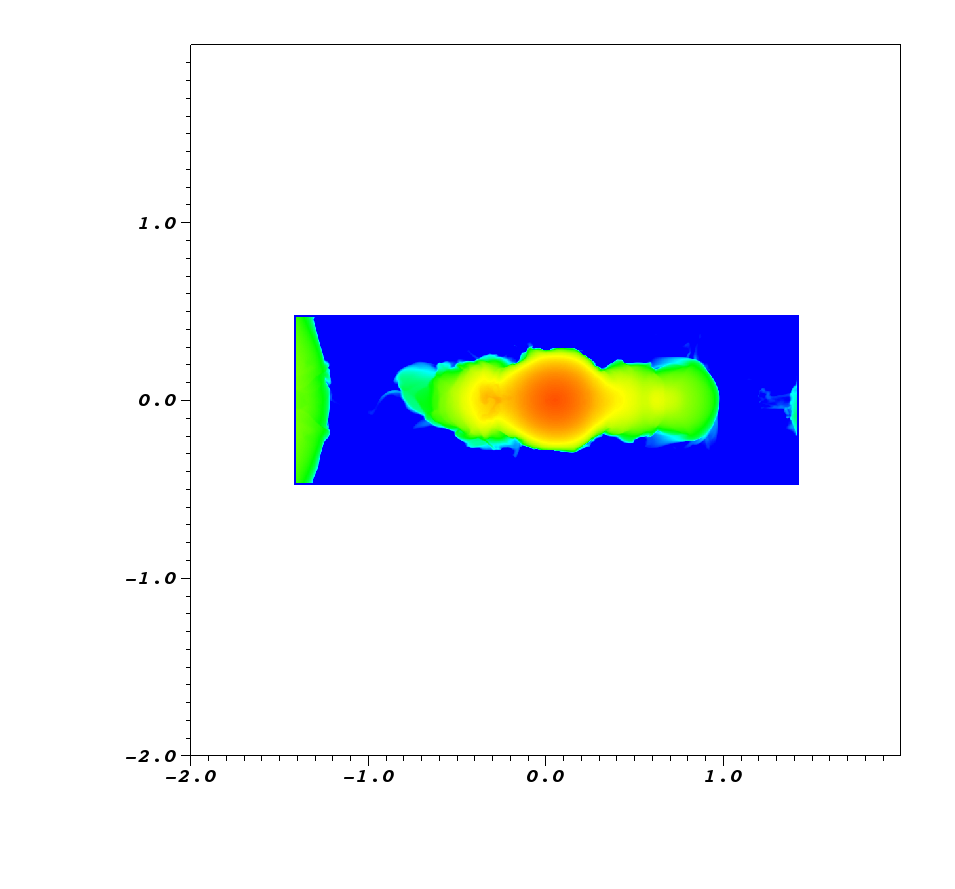}}{  (b) cylHR}
  \vspace{0cm}
  \hspace{0cm}%
\stackunder[-2pt]{\includegraphics[width=8.6cm]{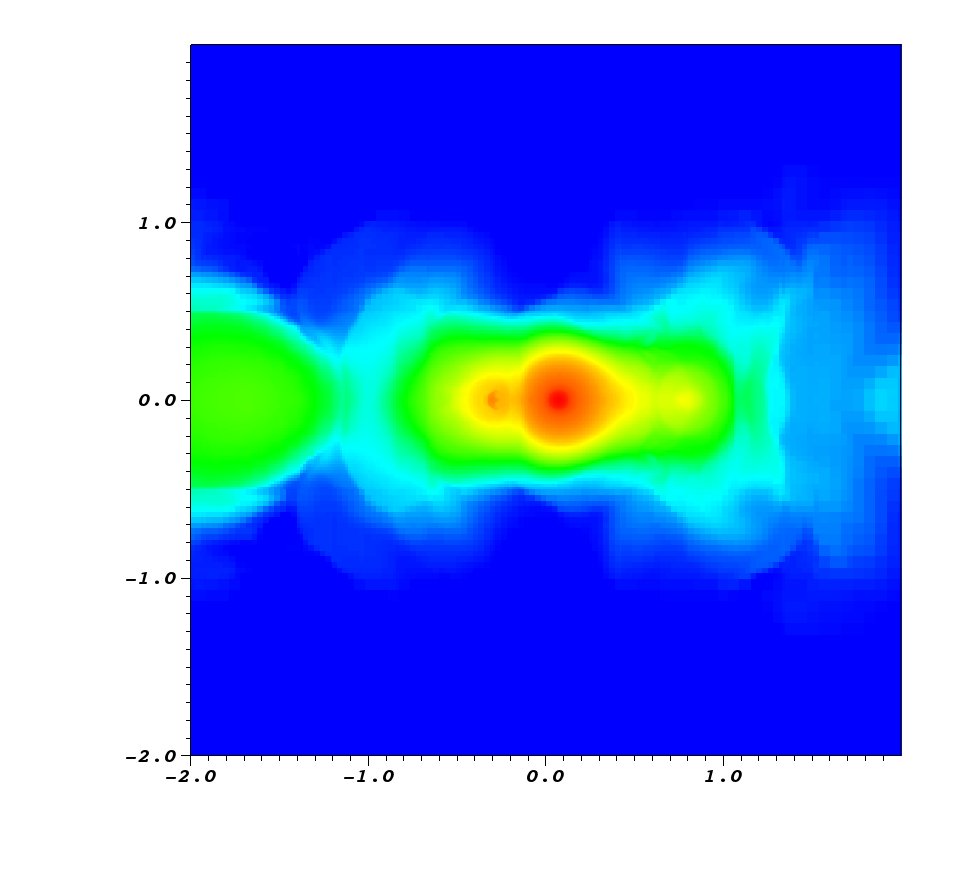}}{  (c) octLR}%
  \hspace{0cm}%
\stackunder[-2pt]{\includegraphics[width=8.6cm]{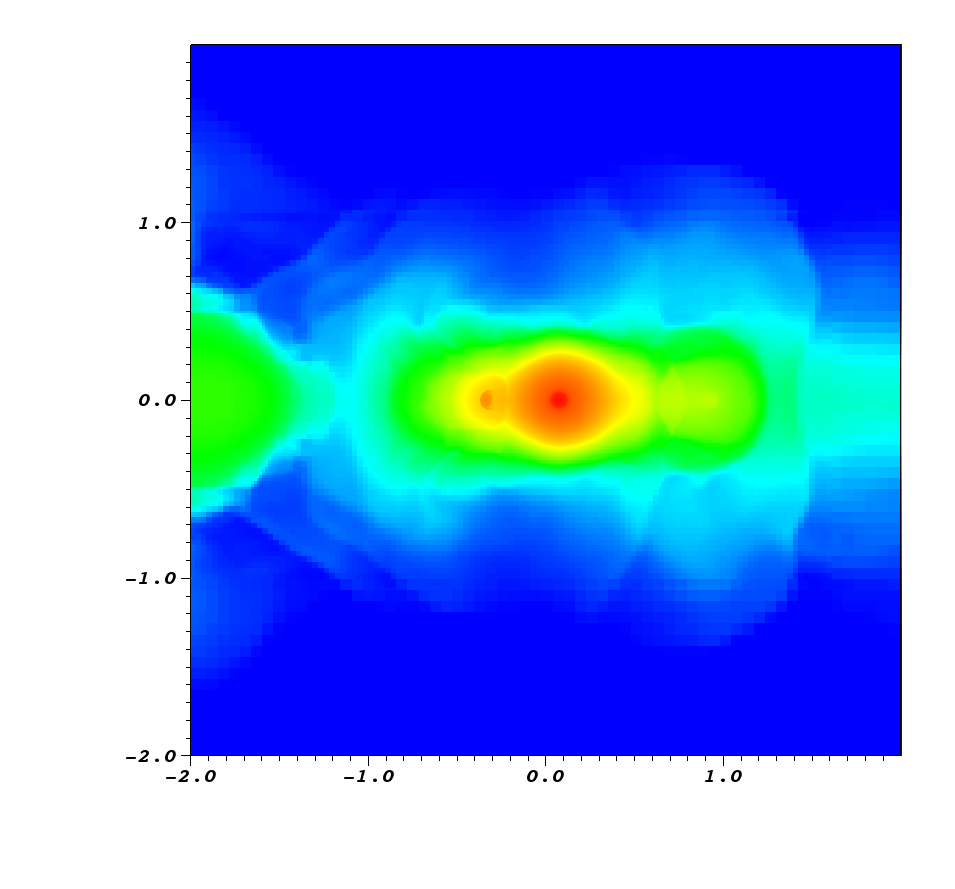}}{ (d) octHR}
\caption{Meridional cross section of the density field at ${ \rm t_{merge} + 0.4}$ or $(t+t_{\rm zpt})/P_0= 24.98$, right after the merger. The heated outer atmosphere as well as the remnants of the donor core can be seen in the Octo-tiger frames.}
\label{fig:mtmerge+0p4}
\end{figure*}

Figure \ref{fig:tmerge+0p4} shows the density in the equatorial plane right after the merger when disruption of the donor {was} complete according to the diagnostic plots. 
However, the remnants of the former core of the donor can be seen being wrapped around the accretor as a long red streak of inspiralling material above a density threshold of about 0.4.
This former core material of the donor was poorly resolved in the Flow-ER simulations. 
The two cores did not collide in a head-on manner. The donor was tidally broken apart and, due to its high specific angular momentum, its material dispersed as a thick disk around the final merged object. 
Figure \ref{fig:mtmerge+0p4} shows the meridional cross section of the density after the merger.
The central, non-spherical merged object has the same basic structure in all four simulations. 
The remnant of the secondary core can be seen in cross section of the Octo-tiger simulations, whereas the core was much more diffuse and poorly resolved in the Flow-ER simulations.
The major difference between the Flow-ER and the Octo-tiger simulations {was} the thickness of the disk in the latter case, due to high temperatures caused by the shock heating during the merger. The shock-heated disk was almost twice as thick with Octo-tiger, as compared to the Flow-ER simulations. 
Notice that all the frames in Figures \ref{fig:tmerge-alot}-\ref{fig:mtmerge+0p4} qualitatively agree very well across all four Q0.7 simulations.

\begin{figure*}
  \includegraphics[width=17.6cm]{figs/color2.png}
  \vspace{0cm}
\stackunder[2pt]{\includegraphics[width=8.6cm]{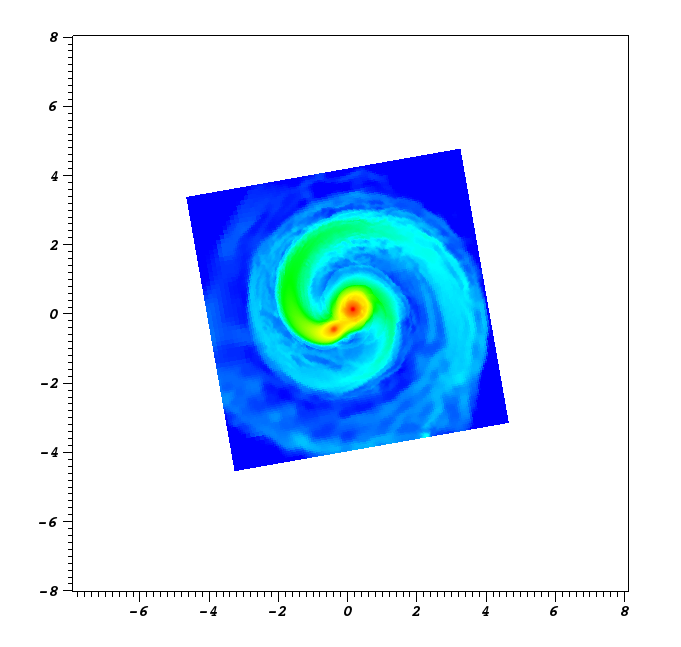}}{(a) octLR}%
\hspace{0cm}%
\stackunder[2pt]{\includegraphics[width=8.6cm]{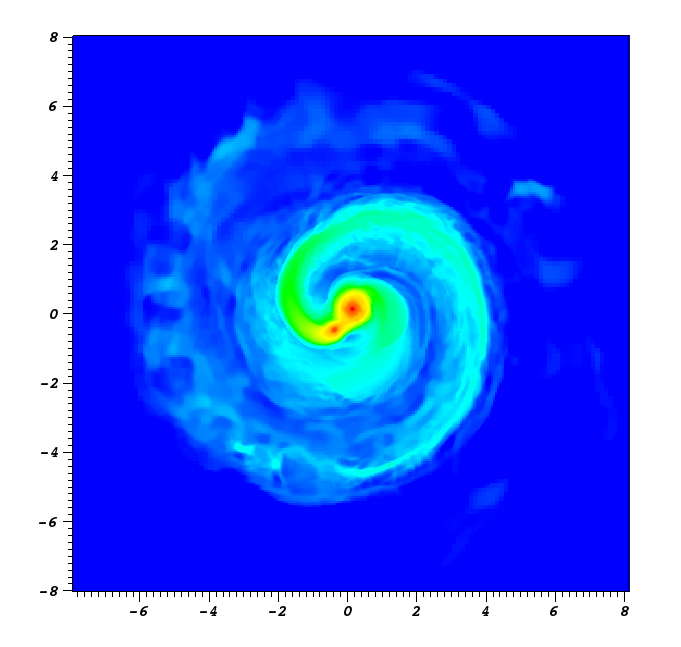}}{(b) octHR}
\caption{Equatorial cross section of the density field at ${ \rm t_{merge} }$ occurring at $(t+t_{\rm zpt})/P_0= 24.58$, showing full octHR grid.}
\label{fig:zotm}
\end{figure*}
\begin{figure*}
  \includegraphics[width=17.6cm]{figs/color2.png}
  \vspace{0cm}
\stackunder[2pt]{\includegraphics[width=8.6cm]{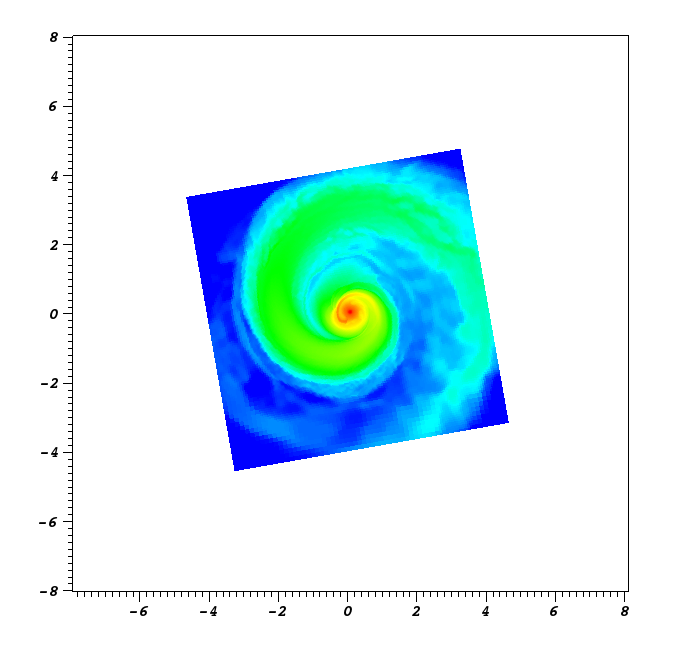}}{(a) octLR}%
\hspace{0cm}%
\stackunder[2pt]{\includegraphics[width=8.6cm]{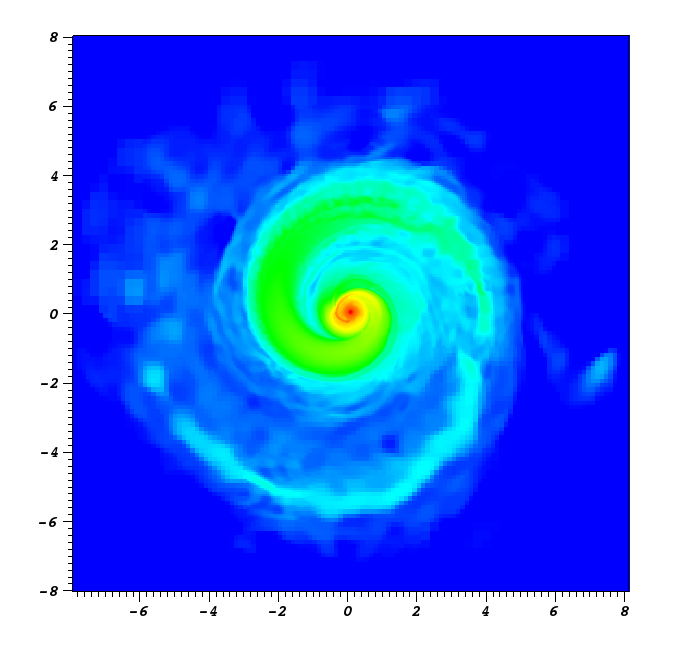}}{(b) octHR}
\caption{Equatorial cross section of the density field at ${ \rm t_{merge} + 0.4}$ or $(t+t_{\rm zpt})/P_0= 24.98$, showing the extended disk and the outflow structure.}
\label{fig:zotm+0.4}
\end{figure*}
\begin{figure*}
  \includegraphics[width=17.6cm]{figs/color2.png}
  \vspace{0cm}
  \hspace{-1.2cm}
\stackunder[-12pt]{\includegraphics[width=9.3cm]{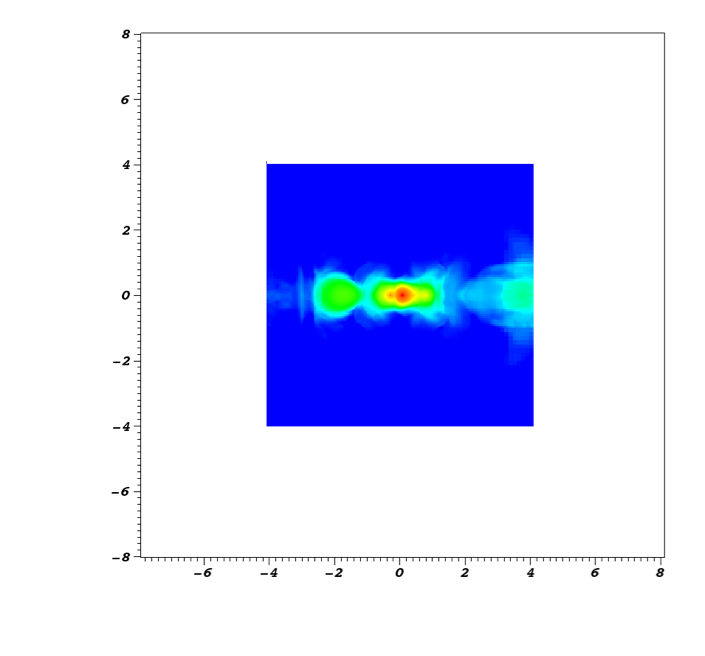}}{(a) octLR}%
\hspace{0cm}%
  \hspace{-0.6cm}
\stackunder[-12pt]{\includegraphics[width=9.3cm]{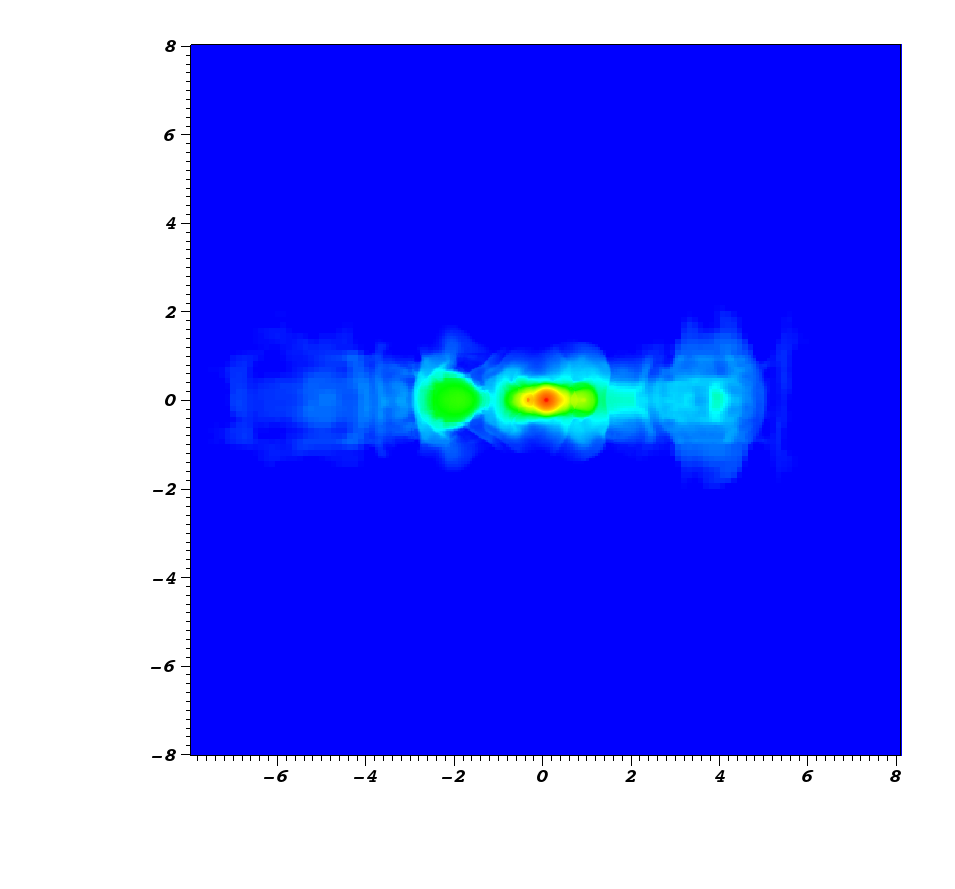}}{(b) octHR}
\caption{Meridional cross section of the density field at ${ \rm t_{merge} + 0.4}$ or $(t+t_{\rm zpt})/P_0= 24.98$. The entire thick disk can be seen in the octHR simulation.}
\label{fig:zotm_side}
\end{figure*}
\begin{figure*}
  \includegraphics[width=17.6cm]{figs/color2.png}
  \vspace{0cm}
\stackunder[2pt]{\includegraphics[width=8.6cm]{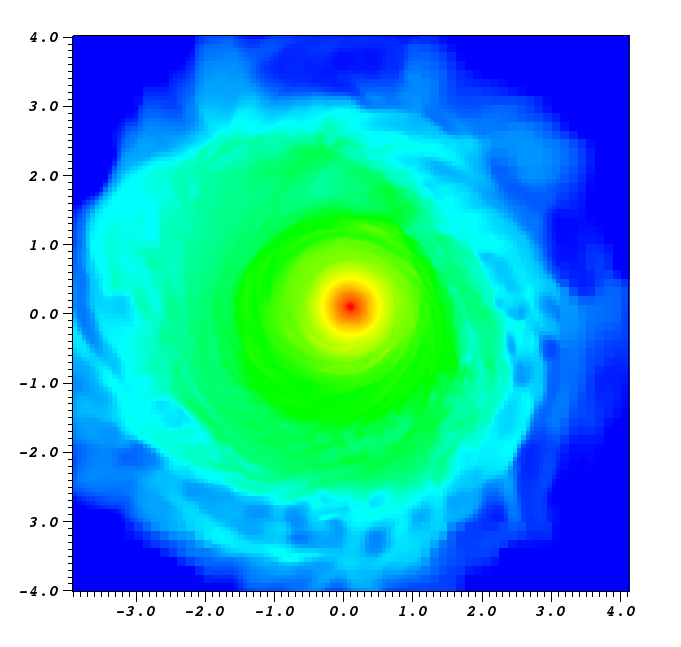}}{(a) Equatorial}%
\hspace{0cm}%
\stackunder[2pt]{\includegraphics[width=8.6cm]{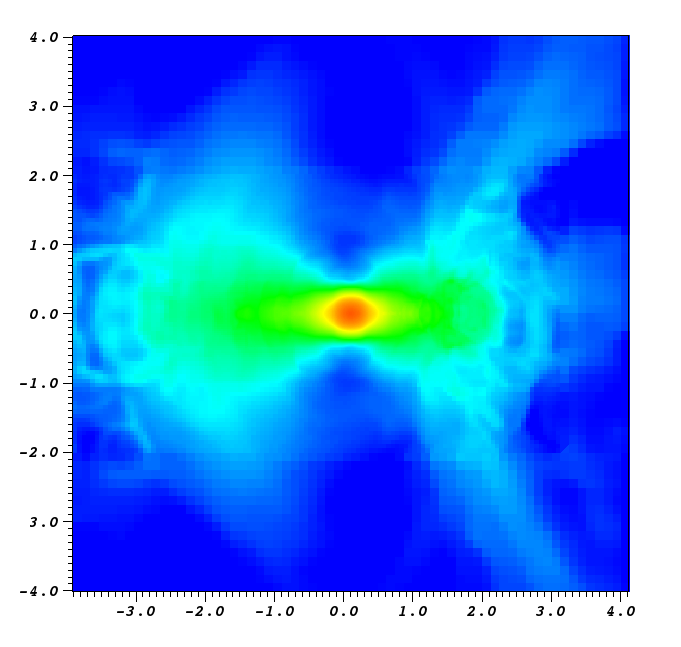}}{(b) Meridional}
\caption{Cross sections of the density field in the octLR simulation at ${ \rm t_{merge} + 3}$. The disk achieved a rough axial symmetry soon after the merger.}
\label{fig:tmerge_long}
\end{figure*}

Figures \ref{fig:zotm} and \ref{fig:zotm+0.4} are identical to Figures \ref{fig:tmerge} and \ref{fig:tmerge+0p4} for Octo-tiger simulations, with the difference that the former pair of frames shows the density slices at the length scale of the entire computational domain of the octHR model. 
The extended, spiral outflow structure, about an order of magnitude larger than the radius of the accretor, was formed in both the Octo-tiger simulations.
A simulation with the Flow-ER code will be computationally very expensive if we achieve similar size of the grid which resolves the circumstellar disk in its entirety, while also trying to maintain enough resolution of the core. 
Figure \ref{fig:zotm_side} shows a meridional cross section of the density field right after the merger, again at the length scale of the octHR simulation. 
The full thick disk is visible in this view in the case of the octHR simulation and most of the material remained in the equatorial plane. 
These frames demonstrate the much better capability of Octo-tiger to resolve the circumbinary disk and the extended outflow structure using AMR. 
Due to computational constraints, we terminated the simulations within about an orbit after the merger. Figure \ref{fig:tmerge_long} shows the density distribution of octLR simulation, which was the only simulation continued, at three orbits after ${ \rm t_{merge}}$. 
Only about 1.6 \% mass and 8.4\% of the total angular momentum of the system was lost at this point.

We now discuss some salient differences among the simulations. 
Figures \ref{fig:rhod} and \ref{fig:rhoa} show plots of the central density of the donor and accretor respectively as the Q0.7 binaries evolve. 
 Consider the evolution of the central density of the Flow-ER simulations first.
 The Flow-ER simulations showed a decrease in the central density of both the stars over a dynamical time scale. 
 As explained in section \ref{sec:flower}, this implies that the cores were not well-resolved. Table \ref{table:simpar} lists the core resolution of the donor and the accretor for the initial Q0.7 models for reference.
We expect this decrease to be a function of resolution, since increased resolution can better resolve the steep density gradients near the core. We can confirm this prediction by comparing the plots for cylLR and cylHR simulations.
For both the donor and the accretor, as the resolution increased, the central density showed less deviation from the initial value of unity. 
Note that the mass transfer began just before the driving phase finished after the first four orbits in each simulation.
In Figures \ref{fig:rhod} and \ref{fig:rhoa} we can observe that as soon as the mass transfer started, the behaviour of the central density changed for both the donor and the accretor in the Flow-ER simulations.
A significant change in the central density of the donor or the accretor did not occur in the case of Octo-tiger simulations before the end of the initial driving phase.
As the mass transfer continued, the central density of the donor in the Octo-tiger simulations decreased, while that of the accretor started to increase. 
The difference between the two Octo-tiger simulations may again be a resolution effect.
The decrease in the central density in the Flow-ER simulations appears to be a major numerical artefact. However, from our comparative analysis of the diagnostic plots as well as the density cross sections we can conclude that this did not significantly affect the dynamical evolution.
The agreement among the results also implies that the simulations with both the hydrodynamic codes were concordant.

\subsection{{Hydrodynamic Evolution of Q0.6 Simulations}}
\label{sec:hydro1}

\begin{figure*}
\begin {center}
\includegraphics [width=14cm] {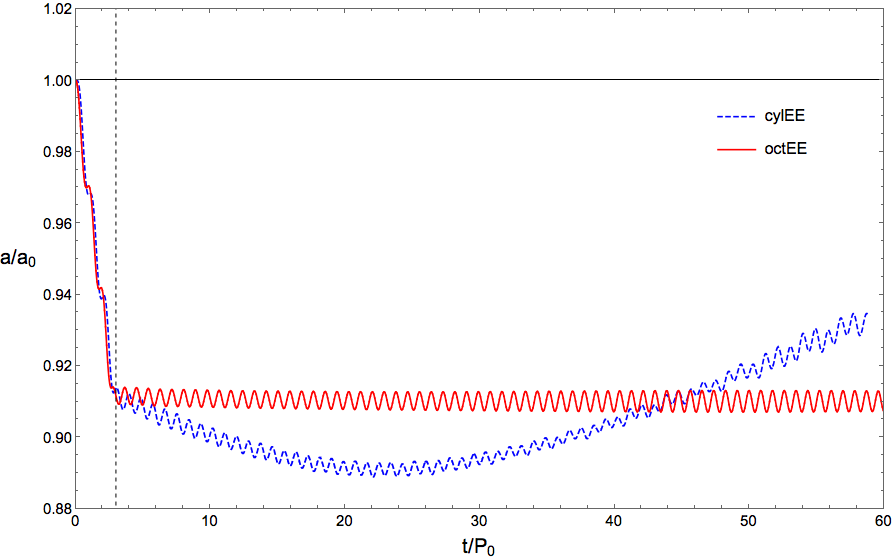}
\end {center}
\caption{Time-dependent behaviour of the normalized separation for Q0.6 simulations. 
The end of the driving phase is marked by the vertical, dashed line. 
Both systems remained separated at the end of the simulations and the differences between the two simulations can be explained on the basis of different mass transfer rates.
}
\label{fig:seplt}
\end{figure*}

\begin{figure*}
\begin {center}
\includegraphics [width=14cm] {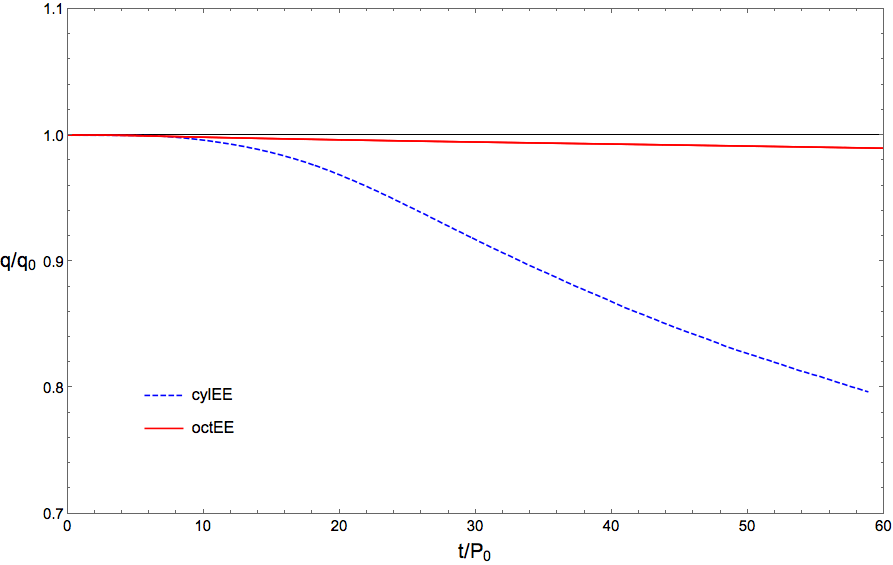}
\end {center}
\caption{Time-dependent behaviour of the normalized mass ratio for Q0.6 simulations. 
The mass ratio of cylEE evolved more rapidly as a result of about 22 times higher mass transfer rate as compared to octEE simulation.
 }
\label{fig:qnormlt}
\end{figure*}

\begin{figure*}
\begin {center}
\includegraphics [width=14cm] {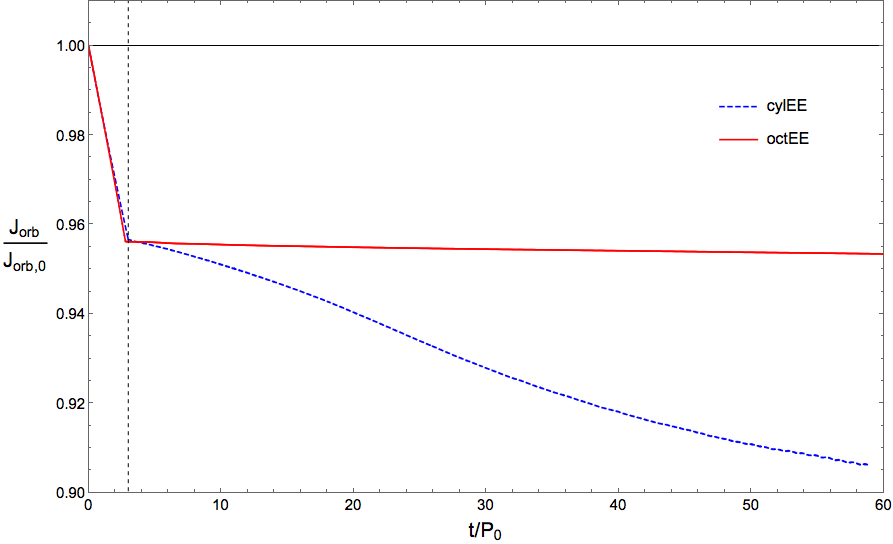}
\end {center}
\caption{Time-dependent behaviour of the normalized orbital angular momentum for Q0.6 simulations. 
The vertical, dashed line marks the end of initial driving phase, which was responsible for the linear drop in the beginning. The rapid orbital angular momentum evolution of cylEE simulation was caused by the higher mass transfer rate.
}
\label{fig:jorblt}
\end{figure*}

\begin{figure*}
\begin {center}
\includegraphics [width=14cm] {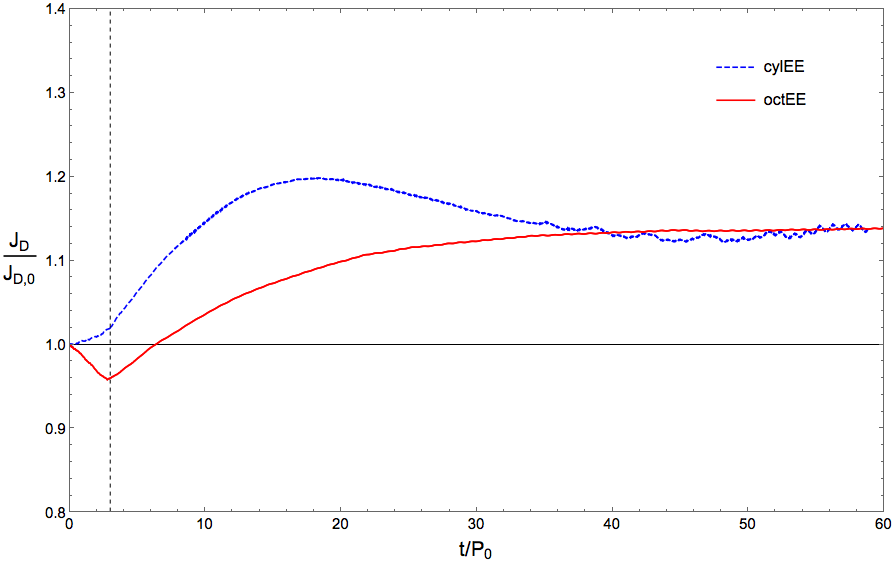}
\end {center}
\caption{Time-dependent behaviour of the normalized donor angular momentum for Q0.6 simulations. The vertical, dashed line marks the end of initial driving phase.
The initial increase in ${ J_{\rm D}/J_{\rm D,0}}$ in the case of cylEE simulation was caused by numerical diffusion.
}
\label{fig:jdlt}
\end{figure*}

\begin{figure*}
\begin {center}
\includegraphics [width=14cm] {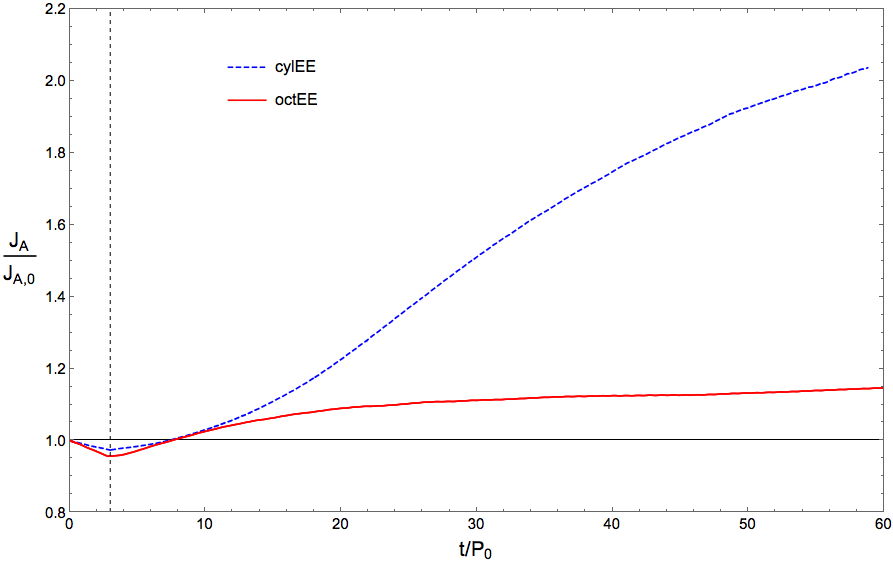}
\end {center}
\caption{Time-dependent behaviour of the normalized accretor angular momentum for Q0.6 simulations. The vertical, dashed line marks the end of initial driving phase.
The accretor angular momentum in the case of cylEE evolved to a higher value as compared to octEE because of numerical diffusion.}
\label{fig:jalt}
\end{figure*}

In this section we describe the evolution of Q0.6 set of simulations- cylEE and octEE. 
As explained in section \ref{tim}, the initial mass ratio lies below the critical value ($q_{\rm crit} \approx 0.63$), hence we expect a stable mass transfer. 
The benchmark for these simulations was to evolve the initial bipolytropic binary for a sufficiently large number of orbits, so that we can demonstrate that the binary systems remain separated, without showing signs of an imminent merger over a dynamical timescale.
These simulations were not repeated at a higher resolution because we have already studied resolution dependance with Q0.7 set of simulations, and it is computationally expensive to run a large number of orbits at a high resolution.
Table \ref{table:simsum2} describes the summary of Q0.6 simulations along with some quantities of interest, and Table \ref{table:movies2} lists the movies of the temporal evolution of the equatorial cross sections of the density fields.
The time-dependent behaviour of key diagnostic parameters is plotted in Figures \ref{fig:seplt}-\ref{fig:jalt}.
We found that Q0.6 binary models were stable on mass transfer with both the codes up to about 60 orbits, and at the end of the simulations the mass transfer rate stabilized to a constant value.
Note that since cylEE simulation was conducted at an overall low resolution, only a general agreement during the evolution can be expected, while the ultimate outcome of achieving a stable mass transfer with both codes is paramount.
After accounting for the resolution limits for the cylEE simulation as well as the difference in the stable mass transfer rates, the diagnostic quantities show a reasonable agreement among the two simulations.

\begin{figure*}
\begin {center}
\includegraphics [width=15cm] {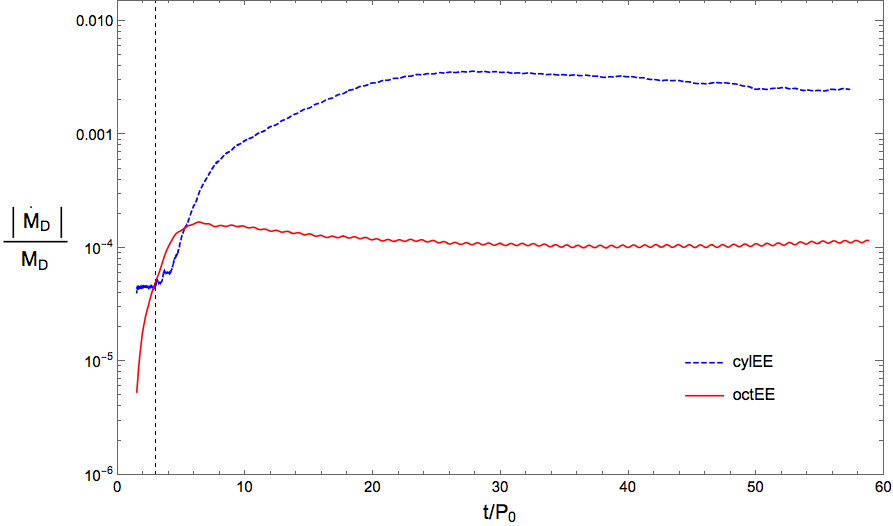}
\end {center}
\caption{Time-dependent behaviour of the smoothed mass loss rate of the donor for Q0.6 simulations. 
The vertical, dashed line marks the end of the initial driving phase. 
The initially increasing $ {|\dot{M}|}/{M_{\rm 0, ref}}$ plots for both simulations ultimately stabilized. The higher rate in cylEE simulation was indirectly caused by numerical diffusion.
}
\label{fig:mdotlt}
\end{figure*}

Consider the the normalized and smoothed mass transfer rate of the donor for Q0.6 set of simulations, plotted in Figure \ref{fig:mdotlt}. The smoothing was done with a moving boxcar average method, identical to that used for Q0.7 simulations.
Each system was driven for 3 orbits at the rate of $1.5\%$ per orbit, and we ensured that a small but steady mass transfer stream was established before the driving was stopped.
The mass transfer rate initially increased and then eventually plateaued for both the simulations. 
Note that this behaviour is qualitatively different from Q0.7 set of simulations.  
Thus as we anticipated, the change in the radius of the donor can keep up with that of its Roche lobe on mass transfer. 
However, the stable normalized mass transfer rate of cylEE simulation, $ {2.5 \times10^{-3}}$, was about 22 times larger than that of octEE simulation, $ {1.1 \times 10^{-4}}$.
We discuss the numerical reasons behind this difference later in this section.
The stable mass transfer rate in real binaries depends on how the angular momentum is removed from the system, as well as the thermal or nuclear evolution of the components. 
The cylEE simulation can be considered to have progressed at a faster rate as compared to octEE, and 
the diagnostic parameters of the former reflected this rapid evolution.

Figure \ref{fig:seplt} shows the normalized separation between the two binary components of Q0.6 set of simulations.
Oscillations with an amplitude of approximately $1\%$ can be noticed throughout the evolution due to small eccentricity of the orbits.
The roughly linear fall in the separation for the first 3 orbits was caused by the artificial removal of angular momentum. 
A total of $4.5\%$ of the angular momentum was removed, resulting in the initial approximately $9\%$ decrease in the separation.
The binary systems remained separated in both the simulations, which is consistent with our analytical expectations.
In the case of cylEE simulation after the driving phase ended, the separation decreased over the next 20 orbits, and then the binary started separating, while the separation for octEE simulation shows little change after the initial driving phase. 
Near the end of the simulations, the cylEE binary also ended up at a larger separation as compared to octEE. 
These differences can be understood on the basis of the different mass transfer rates. 
Due to over an order of magnitude larger stable mass transfer rate, by the end of the simulations, the donor in cylEE transferred about 13\% of its mass to its companion, while the donor in octEE transferred only 0.7\% of its mass. 
This can also be inferred from Figure \ref{fig:qnormlt}, which shows the plot of normalized mass ratio for the Q0.6 simulations.
Since the mass ratio of the Q0.6 binaries was less than unity, the analytical equations predict that the separation should increase on mass transfer.
The timescale over which the separation increases would be proportional to the rate of mass transfer. Thus the change in separation in case of octEE was much smaller than that in cylEE over the length of the simulations. 
The magnitude of the increase in separation would be qualitatively proportional to the amount of mass transferred; thus the binary in cylEE ended up at a greater separation then that in octEE.
At the end of the simulations, the the donor in cylEE was less massive, as compared to the donor in octEE.  
Since the convective envelope with $n_e=3/2$ expands on mass loss, the donor in Q0.6 set of simulations would expand to a larger size.
This expansion of the donor explains the fact that at $t/P_0 \approx 60$, the binary in cylEE ended up at a separation wider than at the beginning of the mass transfer phase, while the donor was still capable of transferring mass.

\begin{table}
	\caption{ Quantities of interest for Q0.6 simulations}
		\label{table:simsum2}
	\begin{center}
	\begin{tabular}{llll} 
		\hline
		 Model ID &  cylEE  &  octEE     \\
		\hline
		{ Equation of state}  &  Ideal gas  & Ideal gas with    \\
		  		         &  & Dual Energy  \\		  
		                        &   & Formalism  \\			 
{Total orbits ($ t/ \rm P_0$)}   & $58.80$  & $ 60.31$  \\
		  Timesteps  &  $ 1.84$ M & $ 1.09$ M   \\
		 {${\rm T_{\rm wall} (d)}$} & $24 $ &  $ 8 $ \\
		{Number of cores}  & $256$  & $ 1280$   \\		  
		 {Cost (CPU-hr/orbit) }            &  $ 2.51 $ k & $ 4.08$ k    \\
		 {Cost (calculations/orbit) }     &  $ 2.10 \times 10^{16}  $  &  $ 2.10 \times 10^{16}  $  \\
		  Machine  &  QueenBee   &  QueenBee    \\
		    &   (LONI) &  (LONI)  \\
		  Processors  & 2.33GHz 4-Core   &   2.33GHz 4-Core   \\
		    & Xeon 64-bit   & Xeon 64-bit   \\
                 \hline
	\end{tabular}
	\end{center}
\end{table}

\begin{table}
	\begin{center}
	\caption{Simulation movies}
		\label{table:movies2}
	\begin{tabular}{lllllllll} 
		\hline
		 Model ID &  {Movies }   \\
		\hline
		 cylEE &  cylEE    \\
		 octEE &   octEE-zoomed   \\
                 \hline
	\end{tabular}
	\end{center}
\end{table}

A comparison of the behaviour of the angular momenta for the Q0.6 simulations requires us to account for small numerical diffusion effects in the  cylEE simulation which are not present in the corresponding octEE simulation.
Figure \ref{fig:jorblt} shows normalized orbital angular momenta for the Q0.6 simulations. 
The initial driving can be noticed in the ${ J_{\rm orb}/J_{\rm orb,0}}$ plot with a linear decline of about 4.5\% over the first three orbits.
Figures \ref{fig:jdlt} and \ref{fig:jalt} show the plots for the normalized spin angular momenta of the donor and accretor respectively. 
Note that all three angular momenta plots for Q0.6 showed changes which are much smaller in magnitude as compared to Q0.7 set of simulations. 
After the end of the driving phase,  $J_{\rm orb}$ decreased and the accretor was spun up in both the cases. 
Figure \ref{fig:jdlt} shows that the donor angular momenta in both the cylEE and octEE simulation levelled out to a similar value.
Thus the spin up of the accretor was at the cost of the orbital angular momentum through the mass transfer.
The change in $ J_{\rm orb}$ as well as $ J_{\rm D}$ in these simulations was qualitatively proportional to the average mass transfer rate of the respective simulation.
However, a sign of discrepancy was seen in the early behaviour of ${J_{\rm D}}$ in cylEE simulation. 
The binary systems in Q0.6 set of simulations were driven for the first three orbits, which removed AM from the system as a whole. 
Thus all three angular momenta -- ${ J_{\rm orb}}$, ${J_{\rm D}}$ and ${ J_{\rm A}}$-- should show an initial decrease before the mass transfer could be established.  
As soon as cylEE simulation started and before any mass transfer could begin, however, the spin angular momentum of the donor started increasing.
This behaviour of ${J_{\rm D}}$ in cylEE simulation was anomalous, since it was neither expected nor seen in the Octo-tiger counterpart. Note, however, that this anomalous increase in ${J_{\rm D}}$ was very small, approximately two orders of magnitude below the angular momentum removed during the driving, and thus unlikely to affect the evolution significantly.

 In previous numerical experiments with single polytropic components the Flow-ER code showed excellent behaviour  \citep{Motl2002, DSouza2006,Even2009}. Thus, most likely
 the early behaviour of $J_{\rm D}$ in cylEE before the onset of mass transfer is due to the numerical diffusion of the bipolytropic cores in the Flow-ER code. We have tested this hypothesis by running three supplementary simulations without driving: 1) Flow-ER with bipolytropic components; 2) Flow-ER with polytropic components; and 3) the same binary with bipolytropic components using Octo-tiger.  Only the first simulation displayed the small anomalous increases in ${J_{\rm D}}$ and ${J_{\rm A}}$ accompanied by diffusion in the cores while the total angular momentum remained constant within code limits. The orbital angular momentum and the separation decreased slightly to compensate for the anomalous increase of the spin angular momenta.
In the interest of conciseness and the primary focus of this paper, we do not present the results of these three simulations in detail.
In spite of the drawbacks of the Flow-ER code, we stress that there was an excellent qualitative agreement between cylEE and octEE simulations. 
We conclude that the simulations with both the hydrodynamic codes were concordant in the case of a stable mass transfer as well over a dynamical timescale.

\begin{figure*}
\includegraphics[width=17.6cm]{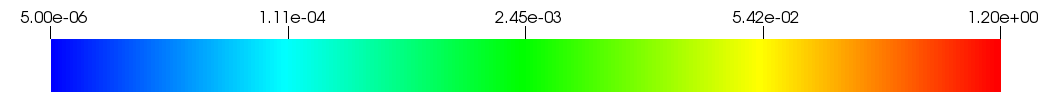}
  \vspace{0cm}
\stackunder[2pt]{\includegraphics[width=8.6cm]{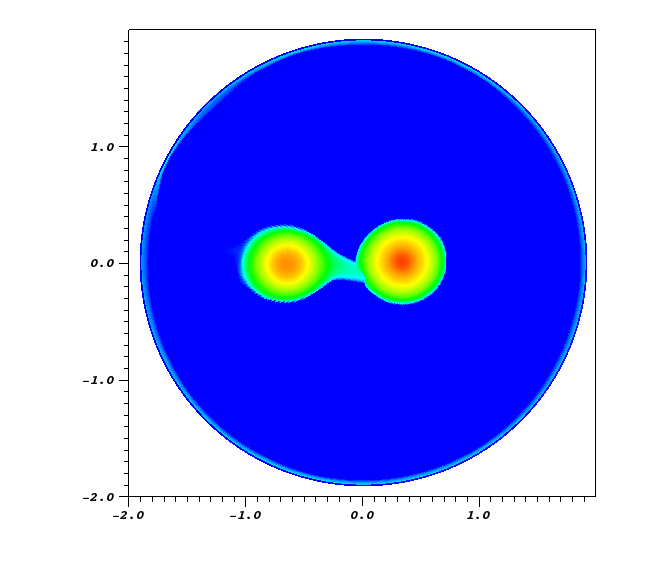}}{(a) cylEE}%
\hspace{0cm}%
\stackunder[2pt]{\includegraphics[width=8.6cm]{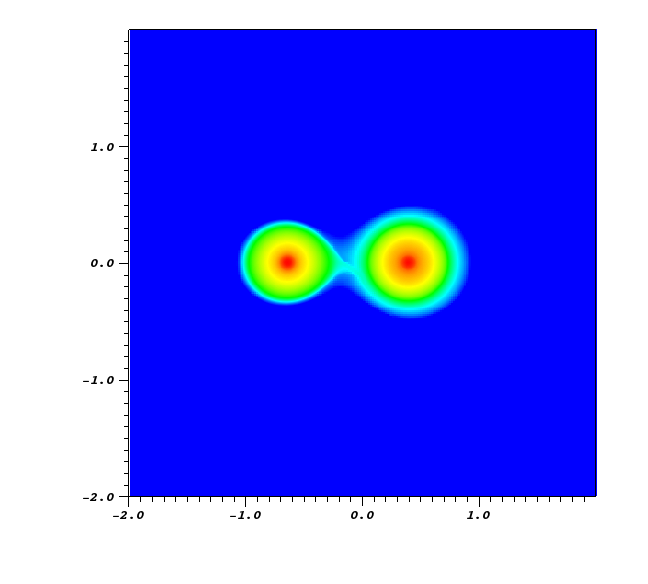}}{(b) octEE}
\caption{Equatorial cross section of the density field at $t/P_0 =45$, when the binary separation was approximately equal for the two Q0.6 simulations.}
\label{fig:q0.6cross}
\end{figure*}

Figure \ref{fig:q0.6cross} shows the equatorial cross sections of the density fields at $t/P_0 =45$, when the binary separation was approximately equal for both the Q0.6 simulations. 
The cross section for cylEE is rotated through a trivial angle for ease of comparison. 
The mass transfer rate in both the simulations was stabilized at this point and a more substantial mass transfer stream can be clearly observed in the case of cylEE. 
As explained in section \ref{sec:hydro}, 
since Octo-tiger code allowed for shock heating in combination with the lack of radiative cooling, 
the components in octEE simulation showed a puffed up atmosphere. 
Note that the minimum value of the density threshold is chosen at a higher value than before in order to make the mass transfer stream more clearly noticeable in octEE simulation. 
At this threshold, the numerical wind in cylEE simulation and most of the puffed up atmosphere in octEE simulation (similar to Figure \ref{fig:tmerge-alot}
 (a) and (c)) is not visible.
The differences between the two codes in terms of the treatment of energy equation did not affect the ultimate outcome of Q0.6 set of simulations.

We calculated the amount of mass and angular momentum lost by the binary system as a whole at the end of Q0.6 simulations. The mass is considered lost if it lied outside the ${ \rm L_2}$ point, and the angular momentum was also measured within this region after accounting for the initial driving phase. 
The binary in cylEE simulation lost a total of about $0.12 \%$ of its mass over the duration of the simulation, however it gained approximately $0.31 \%$ of the total angular momentum. 
These quantities were dominated by two competing numerical effects in Flow-ER simulations. First, there was numerical wind which carried away mass and angular momentum from the binary. Secondly, the implemented density floor reset added a finite amount of mass after each timestep. This mass added angular momentum to the system as well, since the calculations were performed in a rotational frame of reference. Thus cylEE simulation could not accurately track the lost mass and angular momentum over such a large number of orbits. 
The Octo-tiger simulation can give better estimates of the lost mass and angular momentum, since there was no density floor reset or numerical wind.
The binary in octEE simulation lost only about $4.5 \times 10^{-3} \%$ of its total mass, which carried away $0.33 \%$ of the total angular momentum. 
The primary reason for this mass loss through the ${ \rm L_2}$ point was the swelling of the atmosphere of the donor due to the treatment of shock heating. 
The lost material in Octo-tiger simulations can be better tracked with a larger size of the computational domain.

The numerical diffusion and associated artefacts in Flow-ER simulations could be ameliorated by increasing the resolution, however this is computationally prohibitively expensive.
When the grid resolution is doubled for a unigrid simulation, the increase in the computational cost increases by a factor of $2^3$ because of the increase in the size of the data. 
However, since the general benchmark is evolving to a fixed point in time, the \cite{CFL1928} condition needs to be obeyed. Thus the timestep in Cartesian geometry decreases by a factor of two, which makes the computations 16 times more expensive. 
In cylindrical geometry the penalty of Courant-limited timestep is higher, making the computation up to 32 times more expensive in Flow-ER simulations. 
Consider Tables \ref{table:simsum} and \ref{table:simsum2}  where the cost of performing each simulation in terms of computational time is described. 
The cylHR simulation was about 21 times more expensive per orbit, as compared to the cylLR simulation, while the octHR simulation was only about 8 times more expensive than octLR per orbit.
Thus in terms of scaling with increasing problem size, Flow-ER scaled worse than a factor of 16, while Octo-tiger fared much better. 
Octo-tiger can also utilize a much larger number of cores because of its use of the HPX runtime system. 
We would like to emphasize that Octo-tiger code as well as the HPX runtime system both are under active development and are being continuously improved. 
We used a more optimized version of Octo-tiger for octEE simulation, which incorporated adaptive angular velocity during evolution. The binary axis was kept fixed along the X-axis of the computational domain during evolution, thus reducing the need of constant recalculation of the AMR grid (see movie octEE-zoomed listed in Table \ref{table:movies2}). Compared to octLR simulation, which was conducted with a similar effective resolution, octEE simulation was over 8 times more efficient. Thus the code Octo-tiger is much better suited for conducting numerical simulations of close binary interactions as compared to Flow-ER.

\section{{Summary and Discussions}}
\label{summary}

In this paper we established the reliability of the set of grid-based numerical tools -- the BSCF method and Octo-tiger -- for simulating the evolution of bipolytropic binary systems on a dynamical scale, ${ \rm i.e.}$ $ \lesssim 100$ orbits. 
We achieved this through a comparative analysis with an independent and well tested hydrodynamic code Flow-ER. 
With the BSCF method, binary systems with a general core-envelope structure for both the stars can be obtained, which were used as quiet initial conditions for the simulations.
The initial models in the suite of simulations closely matched each other, so that the results can be used as benchmarks.
The codes were compared across binary systems with two mass ratios, $q=0.7$ and $0.6$, the former at two resolutions.
The code Flow-ER evolved a binary adiabatically on a cylindrical mesh with an ideal gas equation of state. Octo-tiger used a dual energy formalism 
 (which allowed for shock heating via solving total energy equation explicitly)
 with ideal gas equation of state and evolved the binary on a Cartesian AMR grid. 
The binary components were initially driven together until a steady accretion stream was formed, and the subsequent dynamical behaviour was analysed without any special assumptions.

We summarize the results obtained from the suite of simulations as follows--
\vspace{-0.1cm}
\begin{enumerate}
\item  Q0.7 set of simulations resulted in an unstable mass transfer and an eventual merger of the systems due to tidal disruption of the donor. This was consistent with our theoretical expectations, since the initial mass ratio was greater than the approximate critical value for such system. 
\item Q0.6 set of simulations resulted in a stable mass transfer, which was observed for about 60 orbits. This behaviour was consistent with our expectations as well, since the initial mass ratio was less than the critical value.
\item In the case of Q0.7 set of simulations, a discrepancy between the two codes was observed with respect to the behaviour of mass transfer rate and angular momenta just before the merger.
This can be accounted for by the higher mass transfer rate in Octo-tiger during the merger due to shock heating, as compared to the Flow-ER simulations.
\item The Flow-ER simulations showed an anomalous decrease in the central densities for both the donor and the accretor, which was caused by numerical diffusion due to relatively poor resolution of the cores. Neither the overall behaviour nor the ultimate fate of the binary systems were affected by this numerical artefact. In the case of Q0.6 set of simulations, indirect effects on the mass transfer rate and angular momenta can be inferred. The effects can be attributed to the change in the moment of inertia of individual components due to the internal redistribution of the mass.
\item The intermediate stages, in terms of diagnostic plots and density cross sections, closely matched for the Q0.7 simulations at two resolutions. The stable behaviour of Q0.6 simulations also reasonably matched for the two simulations. This established the concordance of simulations for both the codes, while also proving that the codes can reliably handle stable as well as unstable mass transfer scenario. 
\end{enumerate}
\vspace{-0.1cm}

Further investigations using Octo-tiger and BSCF method could shed light on the complex evolutionary history of contact binary systems and, especially on dynamical aspects that are not well-understood, such as large-scale circulations, internal structure and stability criteria. 
In particular, in paper III these tools will be used to investigate the merger that led to V1309 Scorpii outburst.
The progenitor contact binary had a rather extreme mass ratio of 0.11, and it is proposed that the \cite{Darwin1880} instability of the progenitor contact binary resulted in the observed merger \citep{Stepien2011V}.
These tools can also be useful in studying the interesting case of KIC 9832227, which may be the first prediction of a luminous red nova event \cite{Molnar2017}. 
With the ability of Octo-tiger to conserve key physical quantities up to machine precision, as well as the increase in efficiency due to the HPX runtime system, we can conduct detailed, high resolution simulations that were previously not feasible.
In conclusion, this study is a step forward in the direction of investigating dynamical interactions of mass transferring binary systems.
In the future, the continuation of this project will allow us to numerically study and better understand the evolution of close and contact binary systems, as well as related astrophysical phenomena.

\section{Acknowledgements}
{We thank the referee, Prof.\ Christopher Tout, for his insightful comments and his suggestion to include Q0.6 set of simulations.}
We wish to acknowledge the support from the National Science Foundation through CREATIV grant AST-1240655. The numerical work was carried out using the computational resources of the Louisiana Optical Network Initiative (LONI) and Louisiana State University's High Performance Computing (LSU HPC).



\bibliographystyle{mnras}
\bibliography{references_file} 








\bsp	
\label{lastpage}
\end{document}